\documentclass[a4paper,11pt]{article}
\usepackage{jheppub} % for details on the use of the package, please
\usepackage[T1]{fontenc} % if needed
\usepackage{ytableau}
\usepackage{mathrsfs}

\preprint{USTC-ICTS/PCFT-25-32}

\title{\boldmath M2-brane indices on Higgs vacua and black holes}

\author{Chiung Hwang,}
\author{Chang Lei,}
\author{and Yuezhang Tang}

\affiliation{Interdisciplinary Center for Theoretical Study, University of Science and Technology of China,\\Hefei, Anhui 230026, China}
\affiliation{Peng Huanwu Center for Fundamental Theory,\\Hefei, Anhui 230026, China}

\emailAdd{chiung@ustc.edu.cn}
\emailAdd{leich@mail.ustc.edu.cn}
\emailAdd{yuezhangtang@ustc.edu.cn}

\abstract{As an exact count of protected states, the superconformal index provides a powerful probe into holography and quantum aspects of gravity, reproducing the Bekenstein--Hawking entropy of supersymmetric AdS black holes in the large-$N$ limit. As a step toward understanding quantum black hole microstates, we study the finite-$N$ index of the 3d ADHM quiver gauge theory, a UV description of the 3d $\mathcal N=8$ SCFT dual to M-theory on AdS$_4 \times S^7$. In this note, we analyze both microcanonical and canonical features of the superconformal index. By computing the index to sufficiently high orders using the factorization formula, we identify signatures of quantum black hole states in the finite-$N$ spectrum of the ADHM quiver, which align with the leading large-$N$ contribution reflecting the holographic dual black hole entropy. Furthermore, we introduce the complex-$\beta$ phase diagram of the index, which exhibits distinct peaks potentially associated with different gravitational saddles. We also examine the Hilbert series limit of the factorized index. Our results demonstrate that the finite-$N$ index encodes rich information about black hole microstates and their quantum gravitational interpretation.}

\begin{document} 
\maketitle
\flushbottom

\section{Introduction}
\label{sec:intro}

The superconformal index \cite{Romelsberger:2005eg,Kinney:2005ej,Bhattacharya:2008zy} has become a powerful tool in the study of holography, providing an exact count of protected states in strongly coupled superconformal theories (SCFTs) and offering a unique window into quantum aspects of their gravitational duals. Remarkably, in many holographic examples, the superconformal index reproduces the Bekenstein--Hawking entropy of supersymmetric AdS black holes in the large-$N$ limit,\footnote{See, for instance, \cite{Cabo-Bizet:2018ehj,Choi:2018hmj,Benini:2018mlo,Choi:2019miv,Choi:2019zpz} for early results in each spacetime dimension. Also see a review \cite{Zaffaroni:2019dhb} for more examples including other types of supersymmetric indices, including the pioneering work on the topologically twisted index \cite{Benini:2015eyy}.} establishing a direct bridge between field-theoretic state counting and semiclassical gravity. While much of the recent progress has focused on 4d $\mathcal N=4$ super Yang–Mills (SYM) theory, the 3d $\mathcal N=8$ SCFT, which is holographically dual to M-theory on AdS$_4 \times S^7$, provides an equally rich setting to explore these ideas.

However, while the large-$N$ limit reveals the thermodynamic behavior of black holes, it conceals the detailed structure of individual microstates. This limitation has motivated growing interest in studying theories at finite $N$, where one can hope to resolve the microstate spectrum directly from the field theory. For example, in the case of 4d $\mathcal N=4$ SYM theory, it was found that even at small $N$, the growth of the (signed) BPS state degeneracies computed from the index closely approximates the black hole entropy inferred from the large-$N$ analysis \cite{Murthy:2020scj,Agarwal:2020zwm}. This suggests that a meaningful notion of black hole microstates persists at finite $N$, and tools such as supercharge cohomology have been employed to identify and construct black hole-like states explicitly at finite $N$ \cite{Chang:2024zqi,Chang:2022mjp,Choi:2022caq,Choi:2023znd,Choi:2023vdm,deMelloKoch:2024pcs,Gadde:2025yoa}.

These developments provide compelling evidence that the finite-$N$ index can serve as a refined probe of quantum gravity. In this note, we therefore aim to examine both the microcanonical and canonical aspects of the \emph{finite}-$N$ superconformal index for the 3d ADHM quiver gauge theory, one of the UV gauge descriptions of the 3d $\mathcal N=8$ SCFT. Our study follows two complementary approaches. First, we compare the signed degeneracies obtained from the finite-$N$ indices for the ADHM quiver with the large-$N$ entropy derived in \cite{Choi:2019zpz}. Despite the small values of $N$, we observe remarkably good agreement between these quantities. Second, regarding it as a canonical partition function, we explore the possible phases of the superconformal index by varying a chemical potential over the complex plane. For standard thermal partition functions, the imaginary part of the chemical potential typically reflects the non-unitarity, such as decay rates or non-conserved particle numbers in open systems. In contrast, for the index, the imaginary part of the chemical potential plays a slightly different role: it interferes with the cancellations between bosonic and fermionic contributions due to their relative sign in the definition of the index. In this sense, the complex chemical potential controls which contributions are amplified or suppressed in the index.

More precisely, for the contribution arising from a conjugate pair of complex saddles, labeled by $a$, their microcanonical signs for given charges, collectively denoted as $Q$, are determined by the imaginary part of the complexified entropy $S_a$ as follows:
\begin{align}
I_{a} (Q) = e^{S_{a}(Q)}+e^{S_{a}^*(Q)} \sim e^{\mathrm{Re} \, S_{a}(Q)} \cos\left(\mathrm{Im} \, S_{a}(Q)\right) ,
\end{align}
where the overall sign is governed by the cosine factor involving $\mathrm{Im} \, S_a(Q)$. The index in the canonical ensemble is then given by
\begin{align}
\mathcal I(\beta) = \sum_a \sum_Q I_a(Q) e^{-\beta Q} \,,
\end{align}
and the contribution of saddle $a$ is maximized when
\begin{align}
\pm\frac{\Delta \mathrm{Im} \, S_a(Q)}{\Delta Q}-\mathrm{Im} \, \beta  = 2 \pi k \,, \qquad k \in \mathbb Z \,,
\end{align}
which determines the location of the peak associated with saddle $a$ in the complex-$\beta$ phase diagram of the index. This analysis suggests that different saddles, regardless of whether dominant or not, can be distinguished by examining the complex-$\beta$ phase diagram.

For this purpose, however, we require a computational tool capable of evaluating the index to sufficiently high orders, which is challenging using the standard matrix integral formula derived from supersymmetric localization \cite{Kim:2009wb,Imamura:2011su}. To overcome this difficulty, we employ the so-called factorized index derived in \cite{Choi:2019zpz}, which proves to be more efficient than the matrix integral formula for numerical computations for the index. Moreover, this formula naturally decomposes the index into contributions from discrete Higgs vacua of the theory with generic real masses. These vacua can also be related to solutions of the Bethe ansatz equations (BAEs) associated with 3d supersymmetric partition functions, reflecting the universality of the factorization structure for partition functions on manifolds that are $S^1$ fibered over $S^2$ \cite{Pasquetti:2011fj,Beem:2012mb,Hwang:2012jh,Taki:2013opa,Fujitsuka:2013fga,Benini:2013yva,Benini:2015noa,Choi:2019dfu,Colombo:2024mts}. This connection to BAE solutions may, in turn, suggest a gravitational interpretation of the saddles encoded in the complex-$\beta$ phase diagram.

This formula has been derived only in the single-flavor case. In this work, we extend the derivation and present a general formula for the ADHM theory with multiple flavors, dual to M-theory on AdS$_4 \times S^7/\mathbb Z_F$, where $F$ denotes the number of flavors. We further explore a particular limit of the factorized index that yields the refined Higgs branch Hilbert series \cite{Razamat:2014pta}, offering insight into the types of states contributing to each Higgs vacuum.

The rest of this note is organized as follows. In Section~\ref{sec:SCI}, we evaluate the superconformal index of the ADHM quiver with a single flavor for various gauge ranks $N$, using the factorization formula. Based on this numerical data, we analyze the complex-$\beta$ phase diagrams for $N = 2$ and $N = 3$, which display expected peaks associated with standard graviton and black hole states. Interestingly, the $N = 3$ case also exhibits an additional peak, suggesting the contribution of an extra quantum gravitational saddle. In Section~\ref{sec:Higgs}, we analyze the index for each Higgs vacuum and find that the additional peak is associated with the Higgs vacua corresponding to linear Young diagrams, where only one of the two scalars $Y$ and $Z$ in the adjoint hypermultiplet acquires a vacuum expectation value. In Section~\ref{sec:factorization}, we extend the previous derivation of the factorized index for the single-flavor case to theories with multiple flavors. The Higgs branch Hilbert series limit is also discussed. Finally, in Section~\ref{sec:discussion}, we conclude with remarks and discuss limitations. The explicit index data is provided in Appendix~\ref{sec:data}.
\\

%\paragraph{Note added.} This is also a good position for notes added
%after the paper has been written.

%\newpage
\section{Superconformal indices of M2-branes at finite $N$ and complex-$\beta$ phase diagrams}
\label{sec:SCI}

The low-energy dynamics of M2-branes probing $\mathbb C^4$ can be described by the three-dimensional $\mathcal N=8$ SCFT, which allows several dual gauge descriptions. A notable pair, related by the 3d $SL(2,\mathbb Z)$ duality, would be the $U(N) \times U(N)$ ABJM theory with the Chern-Simons level $k = 1$ \cite{Aharony:2008ug} and the $\mathcal N=4$ supersymmetric Yang-Mills theory with one fundamental and one adjoint hypermultiplets, dubbed the ADHM quiver theory (with a single flavor in this case).\footnote{The name ``ADHM'' comes from the fact that the Higgs branch of this theory coincides with the instanton moduli space constructed by the ADHM method \cite{Atiyah:1978ri}.} While only part of the $\mathcal N=8$ supersymmetry is manifest in both descriptions, $\mathcal N=6$ and $\mathcal N=4$, respectively, it is enhanced to $\mathcal N=8$ at the IR fixed point. Similarly to 4d $\mathcal N=4$ SYM, these theories are of great interest in the context of the AdS/CFT correspondence because the 3d $\mathcal N=8$ SCFT is holographically dual to M-theory on AdS$_4 \times S^7$. In particular, the large-$N$ limit of their supersymmetric indices have been discussed in great detail and shown to capture the Bekenstein--Hawking entropy of dual black holes \cite{Benini:2015eyy,Hosseini:2016ume,Choi:2019zpz,Choi:2019dfu,Hosseini:2022vho}.\footnote{See \cite{Nian:2019pxj} for a different supersymmetric background. See also \cite{Zaffaroni:2019dhb} for a comprehensive review.}

Although large‑$N$ analysis certainly helps us understand the statistical aspect of black hole entropy in a holographic framework, it unfortunately reveals little about the individual microstates of the black hole. For this reason, there has been another line of discussion on this issue, approached from the opposite direction: instead of taking the large‑$N$ limit, one attempts to understand dual gravity states, including those of black holes, by analyzing the spectrum of a field theory at fixed $N$. Firstly, it was observed that for 4d $\mathcal N=4$ SYM theory, even with very small values of $N$, the enumeration of the number of states using the superconformal index fits surprisingly well with the growth of the number of states expected in the large-$N$ limit \cite{Murthy:2020scj,Agarwal:2020zwm}. Since this large-$N$ result reproduces the Bekenstein--Hawking entropy of the dual black hole \cite{Cabo-Bizet:2018ehj,Choi:2018hmj,Benini:2018mlo,Aharony:2021zkr,Choi:2021rxi}, it is natural to anticipate that there should be a notion of black hole states even for small $N$. Indeed, it was proposed that the graviton states and the black hole states can be distinguished at finite values of $N$ by examining how the exactness of a state under a chosen supercharge changes when $N$ varies \cite{Chang:2024zqi}, and some examples of black hole states have been explicitly constructed within the supercharge cohomology \cite{Chang:2022mjp,Choi:2022caq,Choi:2023znd,Choi:2023vdm,deMelloKoch:2024pcs,Gadde:2025yoa}. Furthermore, it was also shown that the superconformal index captures other types of gravity solutions, e.g., grey galaxies \cite{Choi:2025lck,Deddo:2025jrg}.

Hence, we expect that the superconformal index at finite $N$ contains rich information about the gravity states, which should work not only for 4d $\mathcal N=4$ SYM but also for general holographic field theories, such as the 3d $\mathcal N=8$ SCFT mentioned earlier. However, compared to the 4d case, even the simple enumeration of the number of states is more cumbersome in 3d due to the contributions from the non-perturbative monopole states, which give rise to infinite summations in the standard matrix integral formula for the index that can be obtained from supersymmetric localization \cite{Kim:2009wb,Imamura:2011su}.

In this section, we aim to circumvent this issue by simplifying the computation by means of the so-called factorization method for the 3d superconformal index \cite{Hwang:2012jh,Hwang:2015wna,Hwang:2018uyj}. As we will show shortly, using the factorized index, we are able to compute the (unrefined) index of the 3d ADHM quiver theory up to reasonably high charges to compare with the analytic result from the large-$N$ analysis.
\\

Let us briefly review factorization of the superconformal index of the ADHM theory. Generally, the superconformal index of a 3d $\mathcal N=2$ supersymmetric gauge theory can be defined as a supersymmetric partition function on $S^2 \times S^1$ with periodic boundary condition along $S^1$ \cite{Bhattacharya:2008zy}:
\begin{align}
\mathcal I \; = \;  \mathrm{Tr} (-1)^F e^{-\beta' \{Q,S\}} e^{-\beta (E+j)} e^{-\sum_i \Delta_i Q_i}
\end{align}
where $E$ is the energy, and $j$ is the angular momentum on $S^2$. Each $Q_i$ is a global symmetry charge commuting with chosen supercharges $Q$ and $S = Q^\dagger$ satisfying
\begin{align}
\{Q,S\} = E-R-j \geq 0
\end{align}
where $R$ is the $U(1)_R$ charge. As usual, only the BPS states annihilated by $Q$ and $S$, which thus satisfy
\begin{align}
E = R+j \,,
\end{align}
contribute to the index, and the index is independent of $\beta'$. Using the supersymmetric localization on the Coulomb branch, one can derive the matrix integral formula \cite{Kim:2009wb,Imamura:2011su}, which, for the ADHM quiver with a single flavor, is given by
\begin{align}
\label{eq:matrix_single}
& \mathcal I(w,z,t,q) = \nonumber \\
& \frac{1}{N!} \sum_{m \in \mathbb Z^N} \oint \left(\prod_{a = 1}^N 
\frac{d s_a}{2 \pi i s_a} w^{m_a} t^{-|m_a|/2} q^{|m_a|/2}\right) \times \nonumber \\
&\left(\prod_{1 \leq a \neq b \leq N} \left(1-s_a s_b^{-1} q^{|m_a-m_b|}\right)\right) \left(\prod_{a = 1}^N \frac{(s_a^{-1} t^{-\frac{1}{2}} q^{\frac{3}{2}+|m_a|};q^2)}{(s_a t^\frac{1}{2} q^{\frac{1}{2}+|m_a|};q^2)} \frac{(s_a t^{-\frac{1}{2}} q^{\frac{3}{2}+|m_a|};q^2)}{(s_a^{-1} t^\frac{1}{2} q^{\frac{1}{2}+|m_a|};q^2)}\right) \times \nonumber \\
&\left(\prod_{a,b = 1}^N \frac{(s_a^{-1} s_b t q^{1+|-m_a+m_b|};q^2) (s_a^{-1} s_b z^{-1} t^{-\frac{1}{2}} q^{\frac{3}{2}+|-m_a+m_b|};q^2) (s_a^{-1} s_b z t^{-\frac{1}{2}} q^{\frac{3}{2}+|-m_a+m_b|};q^2)}{(s_a s_b^{-1} t^{-1} q^{1+|m_a-m_b|};q^2) (s_a s_b^{-1} z t^\frac{1}{2} q^{\frac{1}{2}+|m_a-m_b|};q^2) (s_a s_b^{-1} z^{-1} t^\frac{1}{2} q^{\frac{1}{2}+|m_a-m_b|};q^2)}\right) ,
\end{align}
where we have used the shorthand $(a;q^2)$ for the infinite q-Pochhammer symbol, $(a;q^2)_\infty = \prod_{k = 0}^\infty (1-a q^{2 k})$. $w$ $(= e^{-\Delta_w})$, $z$ $(= e^{-\Delta_z})$, and $t$ $(= e^{-\Delta_t})$ are fugacities for the global symmetries of the ADHM theory with a single flavor, shown in Table~\ref{tab:sym} with the corresponding fugacities indicated in their subscript, except the R-symmetry $U(1)_R$, whose fugacity, up to a shift by angular momentum, is $q = e^{-\beta}$.
\begin{table}[tbp]
\centering
\begin{tabular}{|c|ccccc|}
\hline
& $U(N)_G$ & $U(1)_w$ & $U(1)_z$ & $U(1)_t$ & $U(1)_R$ \\
\hline
$\Phi$ & $\mathbf N$ & 0 & 0 & $1/2$ & $1/2$ \\
$\tilde \Phi$ & $\mathbf{\overline N}$ & 0 & 0 & $1/2$ & $1/2$ \\
$X$ & $\mathrm{adj}$ & 0 & 0 & $-1$ & 1 \\
$Y$ & $\mathrm{adj}$ & 0 & 1 & $1/2$ & $1/2$ \\
$Z$ & $\mathrm{adj}$ & 0 & $-1$ & $1/2$ & $1/2$ \\
$V^\pm$ & $\mathbf 1$ & $\pm1$ & 0 & $-1/2$ & $1/2$ \\
\hline
\end{tabular}
\caption{\label{tab:sym} The symmetries of the ADHM quiver gauge theory with a single flavor and the charges of the matter fields (in the $\mathcal N=2$ notation) and, as examples, the monopole operators of the minimal flux, $V^\pm$.}
\end{table}
These symmetries are actually combined and enhanced to the $Spin(8)_R$ symmetry in the IR. As explained, the localization formula for the 3d index includes the infinite summation over quantized magnetic flux on $S^2$.

Using this formula, the large-$N$ behavior of the index has been examined. The leading contribution to the BPS free energy, which is defined as the negative logarithm of the index, is given by \cite{Choi:2019zpz}\footnote{This expression was derived in the Cardy limit, $\beta\rightarrow 0$. While subleading corrections in the Cardy limit have also been discussed \cite{GonzalezLezcano:2022hcf,Bobev:2022wem}, as we will see, the black hole contribution only dominates in the large-charge sector, which is conjugate to small $\beta$. Thus, for our purpose, it is sufficient to focus on the leading expression \eqref{eq:free0}.}
\begin{gather}
\label{eq:free0}
\mathcal F \; = \; -\log \mathcal I \; \approx \; \pm i \frac{4 \sqrt 2 N^\frac32}{3} \frac{\sqrt{\Delta_1 \Delta_2 \Delta_3 \Delta_4}}{2 \beta} \,, \\
\sum_{i=1}^4 \Delta_i = \pm 2 \pi i + 2 \beta \,,
\end{gather}
where $\Delta_i$'s are combinations of $\Delta_{w,z,t}$ and $\beta$, which we do not display here since eventually we are interested in the unrefined limit where $\Delta_w = \Delta_z = \Delta_t = 0$. One condition imposed on them is the real part of $\sqrt{\Delta_1 \Delta_2 \Delta_3 \Delta_4}$ is taken to be positive. See \cite{Choi:2019zpz} for further details.
\\

In addition to this matrix integral formula, there is another useful expression for the index obtained by the factorization method \cite{Hwang:2012jh,Hwang:2015wna,Hwang:2018uyj}. This method rearranges the residues of the integral together with the flux summation, resulting in the factorized form of the index into the vortex and anti-vortex partition functions on $\mathbb R^2 \times S^1$, which is naturally anticipated from the Higgs branch localization \cite{Fujitsuka:2013fga,Benini:2013yva} as well as the holomorphic block \cite{Pasquetti:2011fj,Beem:2012mb}.

More precisely, the factorized index is written as the summation of the product of the perturbative part and the vortex and anti-vortex parts over discrete Higgs vacua of the theory with generic real masses for global symmetries \cite{Choi:2019zpz}:\footnote{The perturbative part $Z_\text{pert}^{\mathcal Y}$ can also be factorized, where each piece can be interpreted as the contribution of perturbative excitations on $\mathbb R^2 \times S^1$ with appropriate boundary conditions.}
\begin{align}
\label{eq:fact_index_single}
\mathcal I(w,z,t,q) = \sum_{|\mathcal Y| = N} Z_\text{pert}^{{\mathcal Y}}(z,t,q) \, Z_\text{vort}^{{\mathcal Y}}(w,z,t,q) \, Z_\text{vort}^{{\mathcal Y}} (w^{-1},z^{-1},t^{-1},q^{-1})
\end{align}
with
\begin{align}
Z^{\mathcal Y}_\text{pert}(z,t,q) &= \left(\prod_{\mathsf a\neq \mathsf b \in \mathcal Y} \left(1-v_\mathsf a^{-1} v_{\mathsf b}\right)\right) \left(\prod_{\mathsf a \in \mathcal Y} \frac{(v_\mathsf a q^2;q^2)}{(v_\mathsf a^{-1} ;q^2)} \frac{(v_\mathsf a^{-1} (t q)^{-1} q^2;q^2)}{(v_\mathsf a t q;q^2)}\right) \nonumber \\
&\quad \times \left(\prod_{\mathsf a,\mathsf b \in \mathcal Y} \frac{(v_\mathsf a v_\mathsf b^{-1} t q;q^2) (v_\mathsf a v_\mathsf b^{-1} (z t^\frac12 q^\frac12)^{-1} q^2;q^2) (v_\mathsf a v_\mathsf b^{-1} (z^{-1} t^\frac12 q^\frac12)^{-1} q^2;q^2)}{(v_\mathsf a^{-1} v_\mathsf b (t q)^{-1} q^2;q^2) (v_\mathsf a^{-1} v_\mathsf b z t^\frac12 q^\frac12;q^2) (v_\mathsf a^{-1} v_\mathsf b z^{-1} t^\frac12 q^\frac12;q^2)}\right) ,
\end{align}
\begin{align}
\label{eq:vort}
Z_\text{vort}^{\mathcal Y}(w,z,t,q) &= \sum_{k_{\mathsf a}}
\left(w t^\frac{1}{2} q^{-\frac{1}{2}}\right)^{\sum_{\mathsf a \in \mathcal Y} k_{\mathsf a}} \left(\prod_{\mathsf a \in \mathcal Y} \frac{(v_{\mathsf a}^{-1};q^2)_{-k_{\mathsf a}}}{(v_{\mathsf a}^{-1} t^{-1} q;q^2)_{-k_{\mathsf a}}}\right) \nonumber \\
&\quad \times \left(\prod_{\mathsf a \neq \mathsf b \in \mathcal Y} \frac{(v_{\mathsf a}^{-1} v_{\mathsf b} z t^\frac12 q^\frac12;q^2)_{-k_{\mathsf a}+k_{\mathsf b}} (v_{\mathsf a}^{-1} v_{\mathsf b} t^{-1} q;q^2)_{-k_{\mathsf a}+k_{\mathsf b}} }{(v_{\mathsf a}^{-1} v_{\mathsf b};q^2)_{-k_{\mathsf a}+k_{\mathsf b}} (v_{\mathsf a}^{-1} v_{\mathsf b} z t^{-\frac12} q^\frac32;q^2)_{-k_{\mathsf a}+k_{\mathsf b}}}\right)
\end{align}
where the vanishing factors appearing in the expression must be discarded. The $U(N)$ ADHM with a single flavor with generic real masses has discrete Higgs vacua that can be labeled by Young diagrams of size $N$, which are denoted by $\mathcal Y$ in \eqref{eq:fact_index_single}. $v_\mathsf a$ for $\mathsf a \in \mathcal Y$ is given by
\begin{align}
v_\mathsf a = z^{\mathsf i (\mathsf a)-\mathsf j (\mathsf a)} (t q)^{\frac{1}{2} (\mathsf i (\mathsf a)+\mathsf j (\mathsf a)-2)}
\end{align}
where $(\mathsf i (\mathsf a),\mathsf j (\mathsf a))$ is the position of box $\mathsf a$ in the Young diagram $\mathcal Y$. $k_\mathsf a$ is a non-negative integer assigned to each $\mathsf a$ such that these integers are non-decreasing along each row and column of $\mathcal Y$, starting from the initial box at position $(1,1)$. Importantly, it turns out this expression is much more efficient than the matrix integral formula \eqref{eq:matrix_single} in numerically calculating the index. Hence, we attempt to evaluate the indices of M2-branes using this expression, for some finite values of $N$, and compare them with the results from the large-$N$ analysis, capturing the Bekenstein--Hawking black hole entropy \cite{Choi:2018fdc,Choi:2019zpz}.

Despite the fact that the general index includes refined charge information, for computational simplicity, we focus only on the overall scaling of the number of states. Thus, as mentioned, we simplify the index by taking $t, \, z, \, w \rightarrow 1$. The resulting index is
\begin{align}
\mathcal I(\beta) \; = \; \mathrm{Tr} (-1)^F q^{E+j} \; = \; \mathrm{Tr} (-1)^F x^Q \; = \; \sum_{Q = 0}^\infty I(Q) \, x^Q
\end{align}
where $x \equiv q^\frac12 = e^{-\beta/2}$ and $Q \equiv 2 E+2 j$. Remember the global symmetries of the ADHM theory with a single flavor are enhanced to the $Spin(8)_R$ symmetry in the IR, whose Cartan charges, conjugate to $\Delta_i$, relate to the $U(1)_R$ charge as follows:
\begin{align}
R = \frac{1}{2} \left(Q_1+Q_2+Q_3+Q_4\right) ,
\end{align}
where $Q_i$'s are integer-quantized. Roughly speaking, the $Q_i$ count the excitations of the adjoint fields $Y$ and $Z$, as well as those of the monopole operators with positive and negative fluxes, respectively. Thus, the single charge $Q$ acting on the BPS states satisfies
\begin{align}
Q = 2 E+2 j = 2R+4 j = Q_1+Q_2+Q_3+Q_4+4 j \,.
\end{align}
Note that the expression \eqref{eq:fact_index_single} is singular in the unrefined limit, which is anticipated because the real part of $\Delta_i$ is associated with the real mass, and the theory has the continuous vacuum moduli space rather than discrete Higgs vacua if the real masses are turned off. It means there should be zero modes whose contribution to the vortex partition function on $\mathbb R^2 \times S^1$ diverges in the unrefined limit. Nevertheless, after the summation over the Higgs vacua, the final superconformal index does not have such a singularity, and one can safely take the unrefined limit. We will revisit this point in Section~\ref{sec:Higgs}.

Accordingly, the leading part of the large-$N$ free energy \eqref{eq:free0} is simplified to
\begin{gather}
\label{eq:free}
\left.\mathcal F\right|_\text{leading} = \left.-\log \mathcal I\right|_\text{leading} = -i \frac{4 \sqrt2 N^\frac{3}{2}}{3} \frac{\Delta^2}{2 \beta} \,, \qquad \Delta = \frac{2 \pi i +2 \beta}{4}
\end{gather}
where we have assumed $\sqrt{\Delta^4} = -\Delta^2$ because the real part of $\Delta^2$ at the extremum will turn out to be negative.
From now on, for simplicity, we will use $\mathcal F$ to denote only the leading part. By taking the Legendre transformation with respect to $\beta$, which is conjugate to $E+j = Q/2$, we obtain (the negative of) the large-$N$ entropy $S$ as follows:
\begin{gather}
S(Q;N^\frac32) = \frac{\sqrt{2} \pi  N^\frac32}{3} \left(\sqrt{1-\frac{3}{\sqrt{2}} \frac{i Q}{N^\frac32}}-1\right) , \label{eq:S} \\
\beta = -\frac{i \pi}{\sqrt{1-\frac{3}{\sqrt2} \frac{i Q}{N^\frac32}}} \,, \label{eq:beta}
\end{gather}
where we have chosen the solution ensuring both $\mathrm{Re} \, \beta > 0$ and $\mathrm{Re} \, S > 0$ as required for the macroscopic entropy. Its exponential should approximate the coefficient of the index for a given $Q$ in the large-$N$ limit.

\subsection{$N = 1$}

Let us start with $N = 1$. The $N=1$ case is rather special because in this case the theory is free and dual to a pair of a hyper and a twisted hypermultiplets. This relation can be understood as both mirror symmetry \cite{Intriligator:1996ex} and Aharony duality \cite{Aharony:1997gp}.\footnote{Recently, it has been shown that a large class of 3d mirror symmetry, and its $SL(2,\mathbb Z)$ generalization, can be derived from 3d Aharony duality \cite{Bottini:2021vms,Hwang:2021ulb,Comi:2022aqo}. Thus, it is not surprising that these two classes of dualities share common examples.} The equality of their indices has been analytically proved in a few different ways, e.g., using the q-binomial theorem \cite{Krattenthaler:2011da} and using the wall-crossing of the vortex quantum mechanics index \cite{Hwang:2017kmk}.

The crucial point is that these decoupled hypers persist even in the higher $N$ cases, implying that there is no black hole state in the $N=1$ case. According to recent discussions, at finite $N$, one can distinguish the graviton states and the black hole states, or more precisely, the states corresponding to smooth horizonless geometries and the others, by their supercharge cohomology \cite{Chang:2024zqi}.\footnote{Very recently, fortuitous states of ABJ(M) theories with $k > 1$, including the large-$k$ regime, have been discussed \cite{Kim:2025vup,Belin:2025hsg,Behan:2025hbx}. It would be interesting to extend these analyses to the $k=1$ case, which is dual to our ADHM theory, and to compare with our results in order to understand the growth of fortuitous states.} It has been proposed that the smooth horizonless states remain BPS for arbitrary $N$, which is why they are called monotone states, while the black hole states are accidentally BPS for some range of $N$, which is why they are called fortuitous states, instead. Since the states in the $N=1$ case remain decoupled and BPS for arbitrary $N$, all of them are monotone and thus correspond to smooth horizonless states.

\begin{figure}[tbp]
\centering % \begin{center}/\end{center} takes some additional vertical space
\includegraphics[width=.49\textwidth]{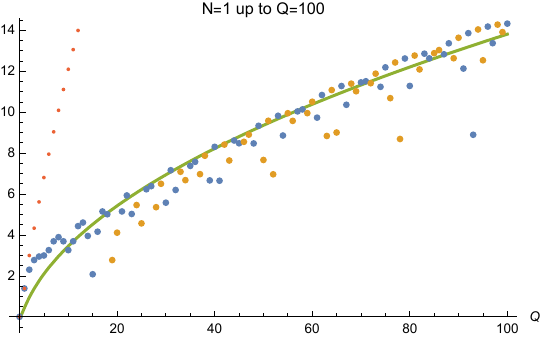}
\hfill
\includegraphics[width=.49\textwidth]{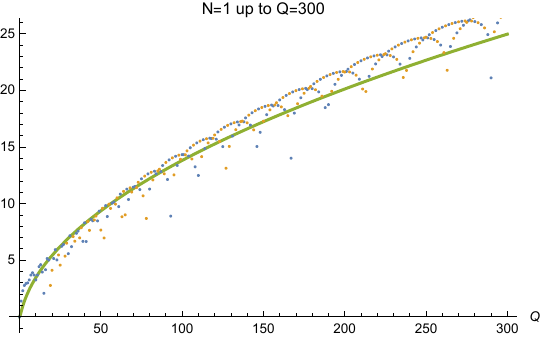}
\caption{\label{fig:n=1} $\log I_{N=1}(Q)$ (blue \& orange dots) vs $S(Q)$ (green line) up to $x^{100}$ on the left and up to $x^{300}$ on the right. Blue and orange dots represent coefficients with positive and negative signs, respectively. In the left plot, the small red dots indicate the graviton spectrum derived from the gravity side. While $\log I_{N=1}$ fits relatively well with $S$ for $Q < 100$, it deviates from $S$ for higher $Q$.}
\end{figure}
Since $N=1$, all the corrections in the large-$N$ expansion are of the same order as the leading contribution. Thus, there is no reason to expect that the actual $N=1$ index and that from the leading large-$N$ free energy \eqref{eq:free} would agree. Furthermore, the fact that the $N=1$ case does not include any black hole state makes the agreement more unlikely because, in the large-$N$ limit, this leading piece is interpreted as the black hole contribution. However, rather unexpectedly, in the actual comparison, this leading contribution provides a reasonably good approximation of the index, at least in the certain range of $Q$. See the left plot in Figure~\ref{fig:n=1}.
For comparison, we also present the graviton index (red dots) derived from the gravity side, which is given by \cite{Bhattacharya:2008zy}
\begin{gather}
\label{eq:graviton}
\mathcal I^\text{graviton}_{N=\infty}(x) \; = \; \mathrm{PE}[g(x)] \; = \; \exp\left[\sum_{k=1}^\infty \frac1k \, g(x^k)\right] \,, \\
g(x) = \frac{(1-x^3)^4}{(1-x)^4 (1-x^4)^2}-\frac{1-x^4+x^8}{(1-x^4)^2} \,.
\end{gather}
The $N=1$ and graviton indices are then evaluated as follows:
\begin{align}
\mathcal I_{N=1}(x) &= 1+4 x+10 x^2+16 x^3+\dots \,, \\
\mathcal I^\text{graviton}_{N=\infty}(x) &= 1+4 x+20 x^2+76 x^3+\dots \,,
\end{align}
which coincide up to $x^1$. We will comment on the meaning of the deviation at higher orders when we examine the indices for $N=2, \, 3$ in the next subsection. Notice that any scalar operator of dimension $1/2$, contributing to the index by $x$, is free according to the superconformal algebra \cite{Minwalla:1997ka}. In the present case, we see $4 x$ in the expansion of the $N=1$ index, which is consistent with the fact that the dual theory is described by a pair of a free hyper and a twisted hyper, or equivalently, four free chirals. For reference, the complete index data can be found in Appendix~\ref{sec:data}.

As anticipated, this agreement with the analytic curve is not permanent. One can reach much higher order of the expansion using the dual free hypermultiplet description, whose index is merely that of four free chirals:
\begin{align}
\mathcal I^\text{dual}_{N=1}(x) \; = \; \exp\left[\sum_{k=1}^\infty \frac1k \, \frac{4 x^k-4 x^{3 k}}{1-x^{4 k}}\right] .
\end{align}
The result is shown in the right plot of Figure~\ref{fig:n=1}, which shows significant deviation at large $Q$. Despite such deviation, we observe that the leading large-$N$ free energy \eqref{eq:free} is good enough to approximate the growth of the $N=1$ index for $Q < 100$, meaning that the states associated with this leading contribution dominates in this charge range. In the following discussions, we focus on the same charge range, which thus allows us to use the large-$N$ expression to approximate the $N=1$ index. This approximation proves useful when examining other examples in the subsequent subsections.

\subsection{$N = 2, \, 3$}

\begin{figure}[tbp]
\centering % \begin{center}/\end{center} takes some additional vertical space
\includegraphics[width=.49\textwidth]{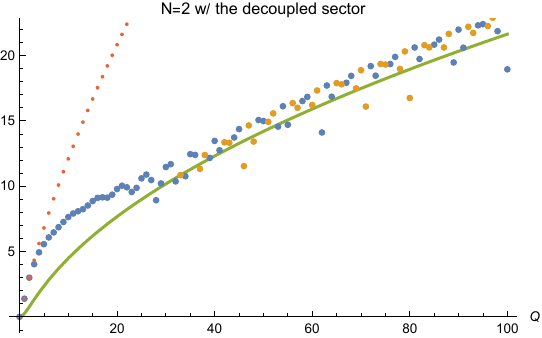}
\hfill
\includegraphics[width=.49\textwidth]{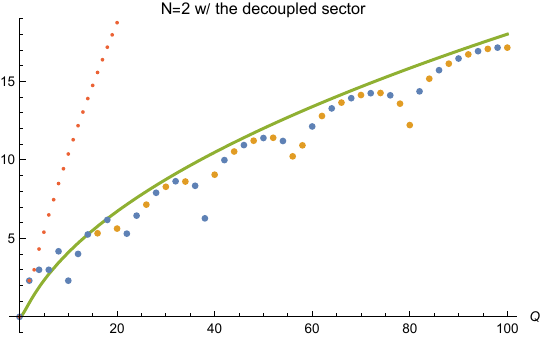} \\
\vspace{0.5cm}
\centering % \begin{center}/\end{center} takes some additional vertical space
\includegraphics[width=.49\textwidth]{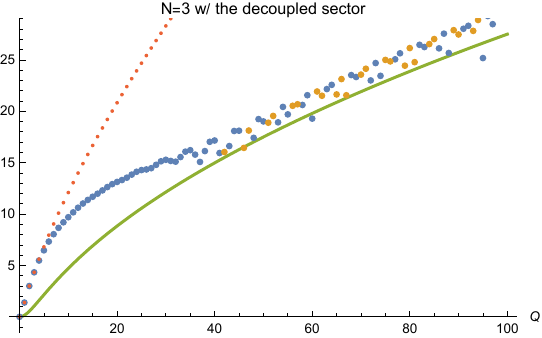}
\hfill
\includegraphics[width=.49\textwidth]{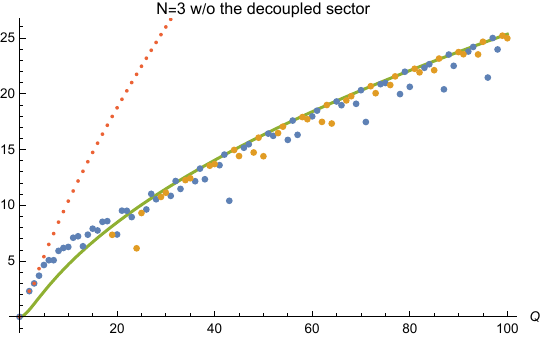}
\caption{\label{fig:n=23} $\log I_{N=2, \, 3}(Q)$ (blue \& orange dots) vs $S(Q)$ (green line) on the left and $\log I_{N=2, \, 3}/I_{N=1}(Q)$ (blue \& orange dots) vs $S_\text{int}(Q)$ (green line) on the right, both up to $x^{100}$. Blue and orange dots represent coefficients with positive and negative signs, respectively, while the small red dots indicate the graviton spectrum derived from the gravity side. The left and right plots in each line differ by the decoupled hypermultiplets, whose index contribution is the same as $I_{N=1}$. }
\end{figure}
Next, we move on to the $N = 2, \, 3$ cases, which are our main examples. Again, one can compute the indices using the factorization formula \eqref{eq:fact_index_single} and compare them with the large-$N$ curves. See the left plots of Figure~\ref{fig:n=23}.

Similarly to the 4d $\mathcal N=4$ SYM case \cite{Murthy:2020scj,Agarwal:2020zwm}, we observe a clear transition, around $x^{20}$ and $x^{30}$ for $N = 2, \, 3$, respectively, after which both the coefficient sign ($+$: blue, $-$: orange) and the logarithm of its absolute value begin to oscillate. Furthermore, the growth of the number of states fits well with the analytic curve (green) obtained from the large-$N$ analysis, up to some overall shift. Thus, it is natural to identify this region as black hole-dominant.

On the other hand, before the transition, where there is no oscillation, we expect this region is dominated by gravitons. For comparison, we again include the graviton index derived from the gravity side in the plots (red dots). The three indices are then given by
\begin{align}
\mathcal I_{N=2}(x) &= 1+4 x+20 x^2+56 x^3+139 x^4+260 x^5+\dots \,, \\
\mathcal I_{N=3}(x) &= 1+4 x+20 x^2+76 x^3+239 x^4+644 x^5+\dots \,, \\
\mathcal I^\text{graviton}_{N=\infty}(x) &= 1+4 x+20 x^2+76 x^3+274 x^4+900 x^5+\dots \,,
\end{align}
where a few lowest order terms of the $N = 2, \, 3$ indices coincide with those of the graviton index. We again see $4 x$ in the index expansion, indicating that there are four decoupled free chirals. For reference, the complete index data can be found in Appendix~\ref{sec:data}.

The fact that the indices agree with that of gravitons only at the lowest few orders even in the graviton-dominant region implies important lessons. From the field theory perspective, the deviation arises due to the trace relation originating from the finite rank of the gauge group, whereas from the gravity perspective, it is attributed to the formation of giant gravitons. The giant gravitons are gravitons carrying large angular momenta in the near horizon geometries, which behave like extended branes \cite{McGreevy:2000cw}; in M-theory on AdS$_4 \times S^7$, they correspond to M5-branes wrapping supersymmetric cycles in $S^7$. Recently, it has been proposed that supersymmetric indices admit expansions in terms of giant graviton contributions as follows \cite{Arai:2020uwd}:
\begin{align}
\frac{Z_N(\Delta_i;\beta)}{Z_\infty(\Delta_i;\beta)} = \sum_{Q_i=0}^\infty e^{-N \Delta_i Q_i} \hat Z_{\{Q_i\}}(\Delta_i;\beta) \,.
\end{align}
One can see that the giant graviton contributions encode the systematic corrections to the graviton index at $N = \infty$ to recover the finite-$N$ index, which are thus the source of the discrepancy between the two indices. As far as we are aware of, the closed form of the giant graviton contributions for the superconformal index of the 3d $\mathcal N = 8$ SCFT is not known yet; partial results instead are available at \cite{Arai:2020uwd,Gaiotto:2021xce,Lee:2022vig,Imamura:2022aua,Hayashi:2024aaf,Beccaria:2023cuo,Hayashi:2025ukc,Deddo:2025lfm,Dorey:2025qht}.
\\

As noted earlier, a theory with $N > 1$ also includes a decoupled pair of a hyper and a twisted hypermultiplets, which should correspond to BPS gravitons. Thus, if we are interested in black hole states, we should look at the interacting sector, whose index can be obtained by dividing its index by that of $N=1$. For comparison with the large-$N$ analytic curve, we should also subtract the contribution of the decoupled sector from the large-$N$ free energy \eqref{eq:free}. Since we have observed that \eqref{eq:free} also explains the growth of the $N=1$ index for $Q < 100$, the logarithm of the index of the interacting sector can be approximated as follows:
\begin{align}
\log \mathcal I_N/\mathcal I_{N=1} \; = \; -\left(\mathcal F_N-\mathcal F_{N=1}\right) \; = \; -i \frac{4 \sqrt2 \left(N^\frac{3}{2}-1\right)}{24} \frac{(\pi i +\beta)^2}{\beta} \,,
\end{align}
where the scaling factor $N^\frac32$ is shifted to $N^\frac32-1$. Thus, the entropy of the interacting sector is obtained by replacing $N^\frac32$ by $N^\frac32-1$:
\begin{align}
\label{eq:Sint}
S_\text{int} (Q;N^\frac32) = S(Q;N^\frac32-1) ,
\end{align}
where $S$ on the right hand side is given by \eqref{eq:S}.
In this way, we can compare the logarithm of the exact indices and the large-$N$ entropy for the interacting sector. The results for $N = 2, \, 3$ are shown in the right plots of Figure~\ref{fig:n=23}. Notice that the agreement between the two becomes more precise; in particular, the transition point appears earlier than the original index including the decoupled sector. This supports the interpretation that the decoupled sector contains only graviton states; by dividing the index by that of the decoupled sector, i.e., the $N=1$ index, we remove part of the graviton contribution, making the black hole contribution more manifest in the resulting index.

\begin{figure}[tbp]
\centering % \begin{center}/\end{center} takes some additional vertical space
\includegraphics[width=.6\textwidth]{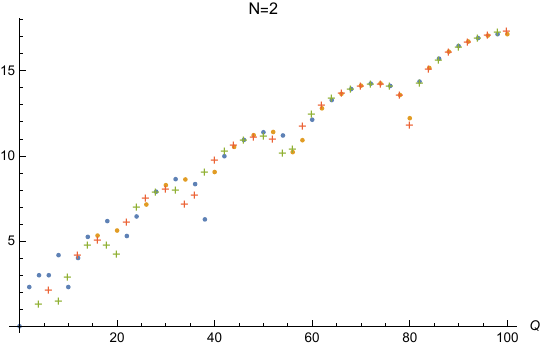}
\caption{\label{fig:n=2-im} $\log I_{N=2}/I_{N=1}(Q)$ (blue \& orange dots) vs $\tilde S(Q;\gamma = 0.53, \, \delta = -0.4)$ (green \& red crosses) up to $x^{100}$. Blue dots and green crosses correspond to coefficients with the positive sign, whereas orange dots and red crosses correspond to coefficients with the negative sign. There is good agreement between the dots and the crosses in the oscillating region for large $Q$.}
\vspace{0.6cm}
\centering
\includegraphics[width=.6\textwidth]{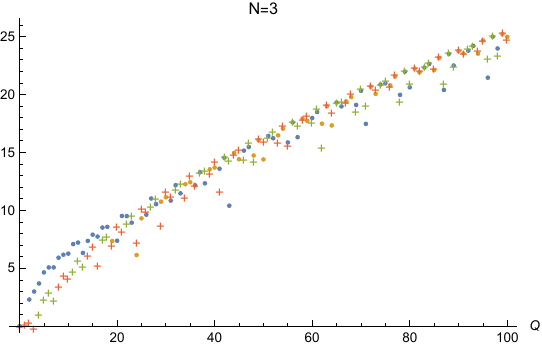}
\caption{\label{fig:n=3-im} $\log I_{N=3}/I_{N=1}(Q)$ (blue \& orange dots) vs $\tilde S(Q;\gamma = 0.95, \, \delta = 0.2)$ (green \& red crosses) up to $x^{100}$. Blue dots and green crosses correspond to coefficients with the positive sign, whereas orange dots and red crosses correspond to coefficients with the negative sign. Other than the points for small $Q$ and those corresponding to $\cos[\dots] \approx 0$ in \eqref{eq:Stilde}, whose logarithm diverges, there is good agreement between the dots and the crosses.}
\end{figure}
Moreover, we observe an interesting sinusoidal pattern from the $N=2$ index after subtracting the decoupled sector contribution. See the top-right plot in Figure~\ref{fig:n=23}. Similar behavior has been observed in the 4d $\mathcal N=4$ SYM case and can be explained by the complex nature of the large-$N$ saddle point \cite{Agarwal:2020zwm}. It turns out our case can also be explained by the same origin. Since a complex saddle must come in pair with its complex conjugate, the actual contribution to the index is $e^{S_\text{int}}+e^{S_\text{int}^*}$, whose logarithm is given by
\begin{align}
\log\left[e^{S_\text{int}}+e^{S_\text{int}^*}\right] \; \approx \; \mathrm{Re} \, S_\text{int}+\log\left|\cos\left[\mathrm{Im} \, S_\text{int}+\dots\right]\right| +\dots \,,
\end{align}
where the oscillation primarily depends on the imaginary part of the complexified entropy. For better visibility of the comparison, we introduce fitted entropy $\tilde S$ with two $O(1)$ parameters $\gamma$ and $\delta$:
\begin{align}
\label{eq:Stilde}
\tilde S(Q;\gamma, \, \delta) = \mathrm{Re} \,S_\text{int}(Q)+\log\left|\cos\left[\mathrm{Im} \, S_\text{int}(Q)+\frac\pi2 (Q-1)-\gamma\right]\right|+\delta
\end{align}
where $\frac\pi2 (Q-1)$ is introduced to reflect the correct period of the sign oscillation of the index. With this minimal modification and some chosen values of $\gamma$ and $\delta$,\footnote{We do not aim to precisely determine the best-fitting values of $\gamma$ and $\delta$, as our goal here is merely to demonstrate that the large-$N$ prediction already provides a good approximation for small $N$. The precise values of $\gamma$ and $\delta$ would become relevant when studying subleading corrections, which have been examined in the Cardy limit \cite{Bobev:2022wem}.} we observe impressive agreement between the exact index and the large-$N$ prediction, unless $Q$ is too small. See Figure~\ref{fig:n=2-im}.
Furthermore, although such a sinusoidal pattern is not manifest in the $N=3$ case, the large-$N$ entropy with the minimal modification also showcases the impressive agreement with the exact index for $N=3$ as well. See Figure~\ref{fig:n=3-im}. Note that most of the deviation arises when $Q$ is small or when the cosine term in \eqref{eq:Stilde} approaches 0, where the logarithm diverges and subleading corrections become significant.
\\

\subsubsection{Complex-$\beta$ phase diagram}

So far, we have focused on the behavior of the index as a function of $Q$, i.e., its microcanonical aspect. However, it is also important to understand the canonical aspect by considering the index as a function of $\beta$. Recall that the key feature distinguishing the black hole-dominant region from the graviton-dominant region is the presence of the sign oscillation, which arises from the definition of the index, where bosonic and fermionic states contribute with opposite signs. Due to this sign difference, it was long believed that the index, as a canonical quantity, could not capture the large entropy associated with black holes. This is true to some extent when $\beta$ is real, but not when $\beta$ is allowed to take complex values. Indeed, this is one of the most important lessons from the large $N$ analysis, which successfully reproduces the black hole entropy assuming complex $\beta$ \cite{Cabo-Bizet:2018ehj,Choi:2018hmj,Benini:2018mlo,Choi:2019zpz}, whereas real $\beta$ leads to the graviton index of $O(1)$ in the large-$N$ limit \cite{Kinney:2005ej,Kim:2009wb}.

Here, we find that complex $\beta$ plays an important role in the finite-$N$ computation as well. First, we introduce the \emph{complex-$\beta$ phase diagram} of the index for the interacting sector by evaluating its absolute value $|\mathcal I_N/\mathcal I_{N=1}(\beta)|$ as $\beta$ varies over the complex plane. We have observed that, obviously, $|\mathcal I_N/\mathcal I_{N=1}(\beta)|$ decreases as $\mathrm{Re} \, \beta$ increases, which is expected since $x = e^{-\beta/2}$, and larger values of $\beta$ suppress more terms in the series expansion. However, richer structure emerges when we compare different values of $\theta \equiv \mathrm{Im} \, \beta/2$.

\begin{figure}[tbp]
\centering % \begin{center}/\end{center} takes some additional vertical space
\includegraphics[width=.31\textwidth]{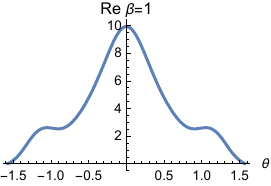}
\hfill
\includegraphics[width=.31\textwidth]{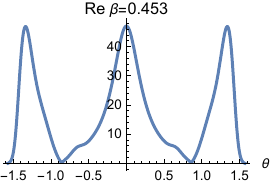}
\hfill
\includegraphics[width=.31\textwidth]{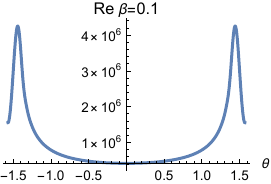}
\caption{\label{fig:n=2-phase} $|\mathcal I_{N=2}/\mathcal I_{N=1}(\beta)|$ as a function of $\theta = \mathrm{Im} \,\beta/2$ when $\mathrm{Re} \,\beta=1$ (left), 0.453 (center), 0.1 (right), respectively. For $N=2$, the periodicity of $\theta$ is $\pi$, half of the other cases because only even powers appear in the series expansion. The transition occurs when $\beta = 0.453$ (center), where there are two peaks, up to sign, one at $\theta = 0$ and the other at $\theta = \pm1.34$.}
\vspace{0.4cm}
\centering % \begin{center}/\end{center} takes some additional vertical space
\includegraphics[width=.31\textwidth]{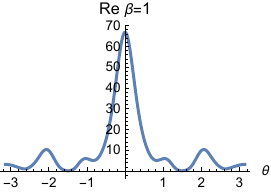}
\hfill
\includegraphics[width=.31\textwidth]{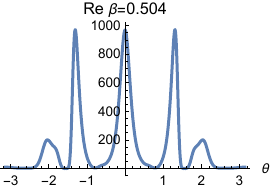}
\hfill
\includegraphics[width=.31\textwidth]{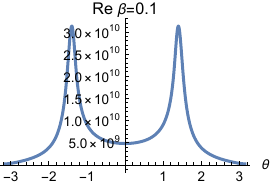}
\caption{\label{fig:n=3-phase} $|\mathcal I_{N=3}/\mathcal I_{N=1}(\beta)|$ as a function of $\theta = \mathrm{Im} \,\beta/2$ when $\mathrm{Re} \,\beta=1$ (left), 0.504 (center), 0.1 (right), respectively. The transition occurs when $\beta = 0.504$, where there are three peaks, up to sign, at $\theta = 0, \, \pm1.31, \, \pm2.04$.}
\end{figure}
Recall the fact that the graviton-dominant region has the uniform sign, while the black hole-dominant region has alternating ones. This means that for nonzero $\theta$, the graviton contributions are diminished due to additional phases that lead to cancellations. In contrast, the black hole contributions will be amplified, as the original cancellations from alternating signs are disrupted. Thus, at large $\beta$, where the index primarily captures small-$Q$ states dominated by gravitons, the absolute value of the index should be maximized at $\theta = 0$. Indeed, we have found that this is the case; see the left plots of Figures~\ref{fig:n=2-phase}~and~\ref{fig:n=3-phase}, which are drawn using the series expansion truncated at $Q = 120$ for better precision.
The two plots show the values of $|\mathcal I_N/\mathcal I_{N=1}(\beta)|$ at fixed $\mathrm{Re} \, \beta = 1$ as a function of $\theta$, the imaginary part of $\beta/2$, for $N=2, \, 3$, respectively. As expected, they exhibit the maximal peak at $\theta = 0$. At small $\beta$, on the other hand, the index mainly captures large-$Q$ states dominated by black hole states, whose contribution should be maximized at nonzero $\theta$. See the right plots of Figures~\ref{fig:n=2-phase}~and~\ref{fig:n=3-phase}, displaying $|\mathcal I_N/\mathcal I_{N=1}(\beta)|$ at fixed $\mathrm{Re} \, \beta = 0.1$. The maximum peaks arise around $\theta = \pm1.44$ for $N=2$ and at $\theta = \pm1.40$ for $N=3$, both of which are nonzero as anticipated.\footnote{The peak values at $\beta = 0.1$ are not saturated when evaluated using the series expansion truncated at $Q = 120$. However, their precise values are not our primary focus and not pursued here; clearly, more accurate results could be obtained by including additional terms in the series. In contrast, the results presented at other values of $\beta$ are all saturated.} Note that the presence of multiple peaks is consistent with the generalized Cardy limit of the superconformal index for the ABJM theory \cite{BenettiGenolini:2023rkq}. In this limit, $\mathrm{Re} \, \beta \rightarrow 0$ while $\mathrm{Im} \, \beta$ is kept finite, and the superconformal index exhibits different growth behavior at large $N$ depending on the value of $\mathrm{Im} \, \beta$. Our results show that such behavior persists at finite $\mathrm{Re} \, \beta$ and finite $N$, i.e., in the non-Cardy quantum regime.

This means that there is a transition at some intermediate value of $\mathrm{Re} \, \beta$, where a peak at nonzero $\theta$ overtakes the peak at $\theta = 0$. This can be interpreted as a type of graviton-black hole transition for the interacting sector, as we expect that a peak at nonzero $\theta$ is primarily associated with the black hole contribution, whereas the peak at $\theta = 0$ is associated with that of gravitons. We have obtained these transition-point values of $\mathrm{Re} \, \beta$ numerically, which are 0.453 for $N = 2$ and 0.504 for $N=3$. $|\mathcal I_N/\mathcal I_{N=1}(\beta)|$ at these points are shown in the center plots of Figures~\ref{fig:n=2-phase}~and~\ref{fig:n=3-phase}, where the peaks at nonzero $\theta$, $\theta = \pm1.34$ for $N = 2$ and $\theta = \pm1.31$ for $N = 3$, match the peak at $\theta = 0$.

Furthermore, for $N=3$, if $\mathrm{Re} \, \beta$ is not too small so that the peak at the first nonzero $\theta$ does not dominate all others, one can observe the crucial fact that there are actually multiple peaks at nonzero $\theta$. This suggests the existence of two distinct collections of states whose sign oscillations have different periods in $Q$. Furthermore, they should be associated with different entropies because, as seen in \eqref{eq:Stilde}, the sign oscillation of the index is governed by the imaginary part of the complex-valued entropy. From $\eqref{eq:Stilde}$, one can estimate the peak value of $\theta$ corresponding to the known black hole entropy. As discussed, the index of the interacting sector is given by
\begin{align}
\mathcal I_N/\mathcal I_{N=1} = \sum_{Q = 0}^\infty I(Q) \, x^Q \,
\end{align}
where $I(Q)$ includes sign. In Figure~\ref{fig:n=3-im}, we have seen that $I(Q)$ can be approximated by the leading large-$N$ entropy
\begin{align}
\label{eq:largeN}
I(Q) \approx e^{\tilde S} = e^{\mathrm{Re} \, S_\text{int}(Q)+\delta} \cos\left[\mathrm{Im} \,S_\text{int}(Q)+\frac\pi2 (Q-1)-\gamma\right] .
\end{align}
In order to minimize the cancellation between different $Q$ so that we have a peak in the phase diagram, the phase difference between adjacent values of $Q$, including that of $x^Q = |x^Q| e^{-i \theta Q}$, should be a multiple of $2 \pi$:
\begin{align}
\pm\left(\frac{\Delta \mathrm{Im} \, S_\text{int}(Q)}{\Delta Q}+\frac{\pi}{2}\right)-\theta = 2 \pi k \,, \qquad k \in \mathbb Z \,.
\end{align}
Given that the period of $\theta$ is $2 \pi$, we simply get
\begin{align}
\label{eq:theta}
\theta = \pm\left(\frac{\Delta \mathrm{Im} \, S_\text{int}(Q)}{\Delta Q}+\frac{\pi}{2}\right) .
\end{align}
Since $\Delta \mathrm{Im} \, S_\text{int}(Q)/\Delta Q$ is not a constant, we need to determine which values of $Q$ dominantly contribute. Recall that, from the Legendre transformation, the relation between $\beta$ and $Q$ is determined as \eqref{eq:beta}, which tells us the dominantly contributing value of $Q$ for a given $\beta$. For example, at the transition-point for $N=3$, $\mathrm{Re} \, \beta = 0.504$, the dominant $Q$ is around $45$, where the least interfering value of $\theta$, determined by \eqref{eq:theta}, is $\pm 1.33$. This value is close enough to the one we obtained from the numerical index data, $\theta = \pm1.31$, given that the latter includes other contributions such as gravitons. Therefore, the peak at $\theta = \pm 1.31$ should correspond to the known black hole solution, whose entropy is governed by $\tilde S$.

\begin{figure}[tbp]
\centering % \begin{center}/\end{center} takes some additional vertical space
\includegraphics[width=.24\textwidth]{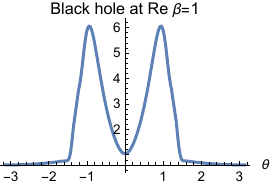}
\hfill
\includegraphics[width=.24\textwidth]{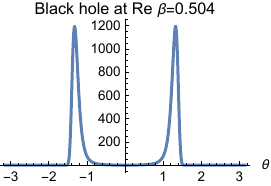}
%\hfill
%\includegraphics[width=.3\textwidth]{bh-phase-0.1.pdf}
\hfill
\includegraphics[width=.24\textwidth]{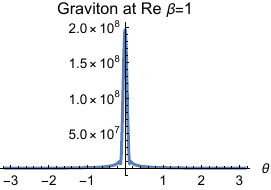}
\hfill
\includegraphics[width=.24\textwidth]{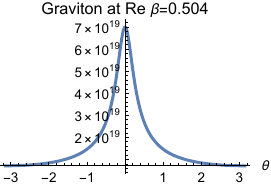}
%\hfill
%\includegraphics[width=.3\textwidth]{graviton-phase-0.1.pdf}
\caption{\label{fig:gravity-phase} The left two are the black hole indices at $\beta = 1$ and $\beta = 0.504$ estimated from the large-$N$ entropy $\tilde S$, given by \eqref{eq:Stilde}. The right two are the graviton indices $\mathcal I^\text{graviton}_{N=\infty}/\mathcal I_{N=1}$ at $\beta = 1$ and $\beta = 0.504$.}
\end{figure}
As a further consistency check, we also plot the complex-$\beta$ phase diagrams for the black hole index governed by $\tilde S$ in \eqref{eq:largeN} and for the graviton index \eqref{eq:graviton}, in place of the $N=3$ index data. As anticipated, the black hole phase diagram exhibits a single peak at nonzero $\theta$ (modulo sign), while the graviton phase diagram shows a peak at $\theta = 0$. See Figure~\ref{fig:gravity-phase}.

Most importantly, the small peak at $\theta = \pm 2.04$ of the center plot in Figure~\ref{fig:n=3-phase} corresponds to neither of these, strongly suggesting that it is associated with a different type of gravitational solution.
While we approximate the index using the large-$N$ entropy $\tilde S$ in \eqref{eq:largeN}, generically, the index would get contributions from multiple saddles, labeled by $a$:
\begin{align}
\mathcal I(\beta) = \sum_{a} \sum_Q I_a(Q) e^{-\beta Q}
\end{align}
with
\begin{align}
I_{a} (Q) = e^{S_{a}(Q)}+e^{S_{a}^*(Q)} \sim e^{\mathrm{Re} \, S_{a}(Q)} \cos \mathrm{Im} \, S_{a}(Q)
\end{align}
where $S_a(Q)$ includes all the corrections, such as $\frac{\pi}{2} (Q-1)$ in the argument of the cosine factor and other $O(1)$ shifts in \eqref{eq:largeN}. The location of the peak in the phase diagram associated with saddle $a$ is then given by
\begin{align}
\theta_a = \pm\frac{\Delta \mathrm{Im} \, S_a(Q)}{\Delta Q} \,.
\end{align}
In other words, our peak at $\theta = \pm2.04$ indicates that there should be a saddle where $\frac{\Delta \mathrm{Im} \, S_a(Q)}{\Delta Q} \approx 2.04$ for dominantly contributing $Q$. Although its precise nature remains unclear, we will see in Section~\ref{sec:Higgs} that this peak is related to a particular Higgs vacuum with vortex excitations, which can also be interpreted as a solution to the Bethe ansatz equations (BAEs) associated with 3d supersymmetric partition functions \cite{Choi:2019dfu}. This may hint at a connection between the peak at $\theta = \pm 2.04$ and a gravitational solution distinct from the well-known graviton gas or black hole configurations, as there is evidence that a particular set of BAE solutions corresponds to the contribution of a specific gravitational background, at least in the unrefined limit \cite{Aharony:2021zkr}.

\begin{figure}
\centering % \begin{center}/\end{center} takes some additional vertical space
\includegraphics[width=.31\textwidth]{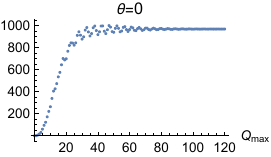}
\hfill
\includegraphics[width=.31\textwidth]{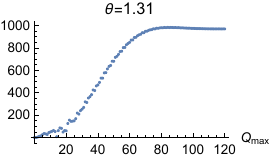}
\hfill
\includegraphics[width=.31\textwidth]{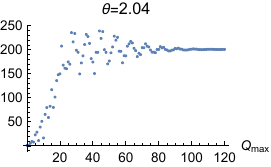}
\caption{\label{fig:n=3-stab} $|\mathcal I_{N=3}/\mathcal I_{N=1}(\beta)|$ for different truncation orders, $Q_\text{max}$, evaluated at the three peaks ($\theta = 0$ (left), $\theta = 1.31$ (center), $\theta = 2.04$ (right)) at the transition point ($\mathrm{Re} \, \beta = 0.504$).}
\end{figure}
We conclude this subsection with one final comment. The series expansion of the index we have used to analyze the complex-$\beta$ phase diagram is truncated at $Q = 120$. Therefore, it is important to verify whether this truncated series includes a sufficient number of terms to reliably estimate $|\mathcal I_N/\mathcal I_{N=1}(\beta)|$, especially at the transition points. To this end, we have evaluate $|\mathcal I_N/\mathcal I_{N=1}(\beta)|$ using various truncation orders in $Q$ and confirmed that keeping terms up to $Q = 120$ yield results sufficiently close to the saturated values, except at $\beta = 0.1$, which requires additional terms in the expansion. Nevertheless, the accurate values at other $\beta$ are enough for our purposes, so we do not pursue further refinement here. As an example, we provide the case for $N = 3$ at the transition point, $\beta = 0.504$ in Figure~\ref{fig:n=3-stab}.

\subsection{$N = 4, \, 5$}

Our last examples are $N = 4, \, 5$. However, these cases require significantly higher computational costs compared to the $N = 2, \, 3$ cases due to larger number of possible Young diagrams, making it difficult to evaluate the indices to sufficiently high orders. Fortunately, our primary interest lies in the overall growth of the index rather than in the precise coefficients, which allows for approximations that reduce the computational burden.

To this end, we first note that the vortex part \eqref{eq:vort} is expressed as a series in $w$, the vorticity fugacity, rather than in $x$. Thus, we need to determine the maximum vorticity contributing to a given power of $x$, say $x^n$, so that we can truncate the vorticity sum at an appropriate order. One can verify that the largest vorticity contributing to $x^n$ is $n$. Namely, to compute the index up to $x^{100}$ as we did in the previous subsections, we require vortex contributions up to vorticity 100.

On the other hand, if we are concerned with the scaling of the overall growth rather than the
precise index, the contributions from high vorticity may be less significant. The
reason is that, as we have seen in the $N = 2, \, 3$ cases, if $Q$ is not too small, the dominant
contribution to the index comes from the black hole states, whose contribution can be approximated
using the large-$N$ free energy \eqref{eq:free0}, which is symmetric under the permutations of $\Delta_i$. Thus, the entropy obtained by
the Legendre transformation will also be symmetric under the permutations of charges $Q_i$, indicating that for large enough charges $Q_i$, the primary contribution will be attributed to equal charge sector. As explained earlier, one of the $Q_i$ counts the monopole flux, or, from the gapped theory perspective, the vortex charge. Thus, for given $Q$, the dominant vorticity would be $Q/4$ if we ignore the angular momentum $j$ for the moment so that $Q \approx Q_1+Q_2+Q_3+Q_4$. Namely, the vortex contributions at vorticity $n/4$ and nearby would be sufficient to capture the overall growth of the index up to $x^n$; and it becomes even lower if we take into account the nonzero angular momentum required by the charge relation for a BPS black hole \cite{Kostelecky:1995ei}.

\begin{figure}[tbp]
\centering % \begin{center}/\end{center} takes some additional vertical space
%\vspace{2cm}
\includegraphics[width=.49\textwidth]{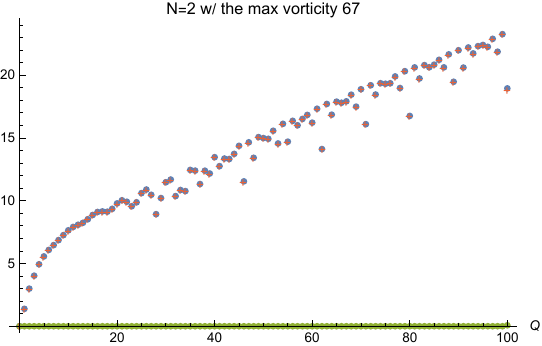}
\hfill
\includegraphics[width=.49\textwidth]{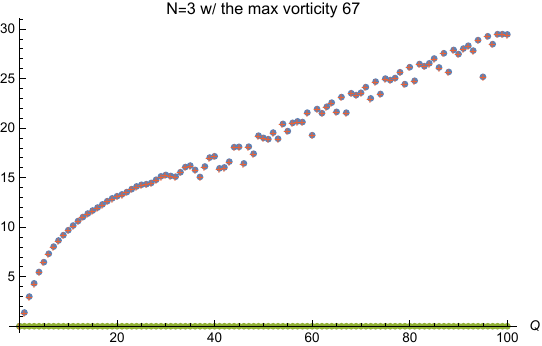} \\
\vspace{0.5cm}
\includegraphics[width=.49\textwidth]{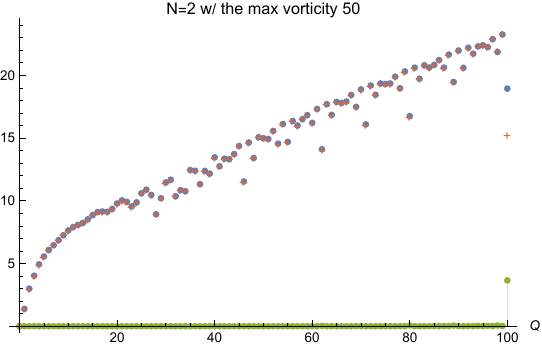}
\hfill
\includegraphics[width=.49\textwidth]{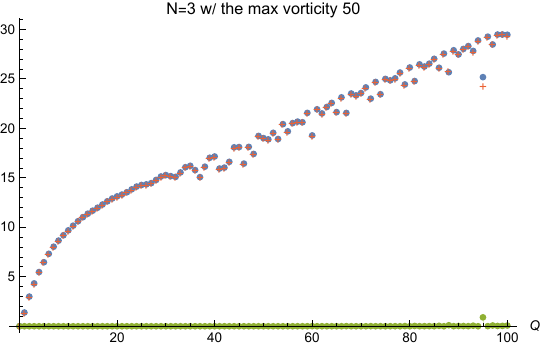} \\
\vspace{0.5cm}
\includegraphics[width=.49\textwidth]{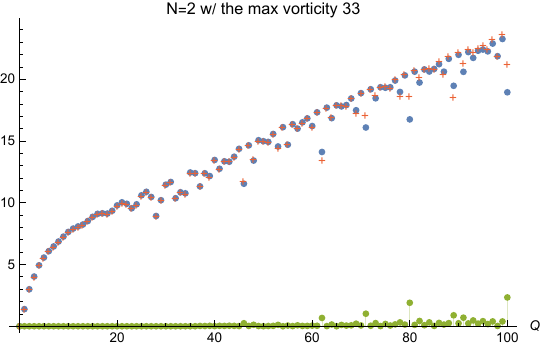}
\hfill
\includegraphics[width=.49\textwidth]{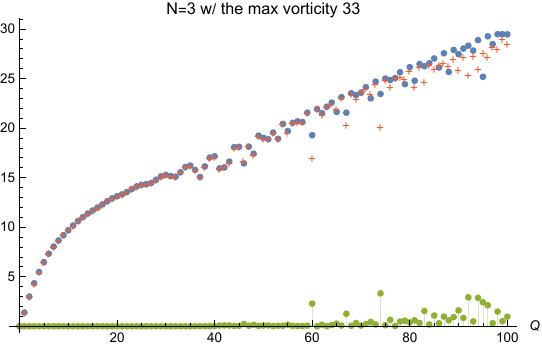} \\
\caption{\label{fig:truncated vortex} Exact $\log I_{N}(Q)$ (blue dots) vs those with truncated vortex contributions (red crosses) for $N=2$ on the left and for $N=3$ on the right. The green dots represent the difference between the exact value and the approximated value with the truncation. We have tested the maximal vorticity of 67, 50, and 33, which are $2/3$, $1/2$, and $1/3$ of the required value when computing the index up to $Q=100$, respectively.}
\end{figure}
Indeed, we have tested the approximation of the $N = 2, \, 3$ indices up to $x^{100}$ by
truncating vortex contributions at different vorticities $67$, $50$, and $33$, corresponding to $2/3$, $1/2$, and
$1/3$ of the maximum vorticity required for the exact computation. The results are shown in Figure~\ref{fig:truncated vortex}.
We observe that the approximation with vorticity up to 67 is nearly indistinguishable from the exact index on a logarithmic scale. The approximation with vorticity up to 50 is also quite accurate, while truncating at 33 results in noticeable deviations, though the overall growth remains qualitatively similar.

\begin{figure}[tbp]
\centering % \begin{center}/\end{center} takes some additional vertical space
\includegraphics[width=.49\textwidth]{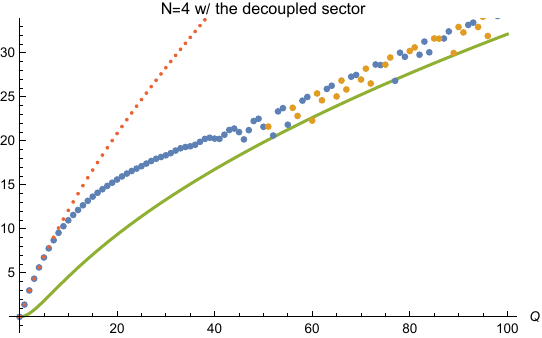}
\hfill
\includegraphics[width=.49\textwidth]{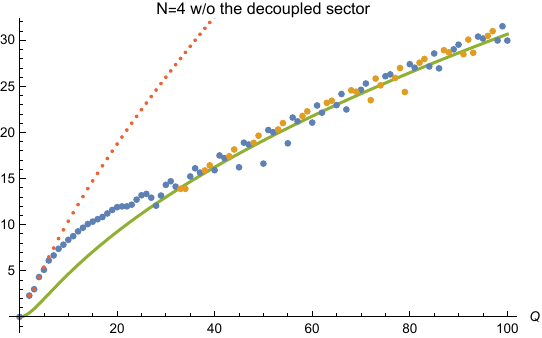} \\
\vspace{0.5cm}
\includegraphics[width=.49\textwidth]{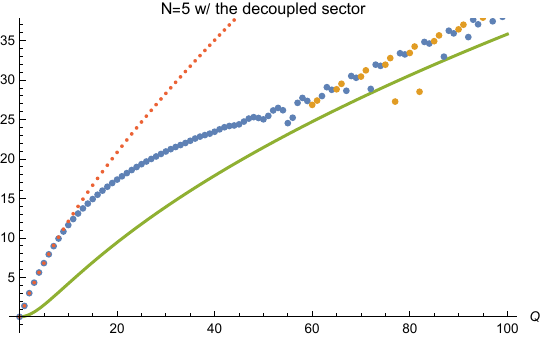}
\hfill
\includegraphics[width=.49\textwidth]{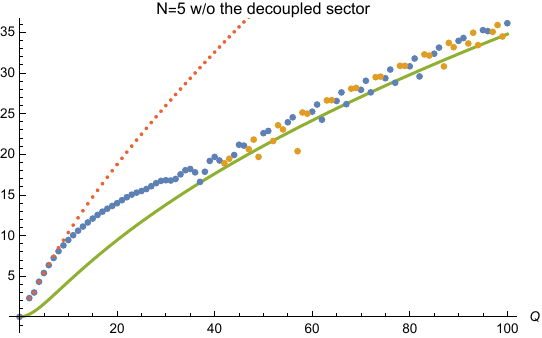}
\caption{\label{fig:n=45} $\log I_{N=4, \, 5}(Q)$ (blue \& orange dots) vs $S(Q)$ (green line) on the left and $\log I_{N=4, \, 5}/I_{N=1}(Q)$ (blue \& orange dots) vs $S_\text{int}(Q)$ (green line) on the right, both up to $x^{100}$. The $I_N(Q)$ used here is an approximate result, obtained by keeping vortex contributions up to vorticity 67 for $N = 4$ and up to 50 for $N = 5$. As before, blue and orange dots represent coefficients with positive and negative signs, respectively, while the small red dots indicate the graviton spectrum derived from the gravity side. The left and right plots in each line differ by the decoupled hypermultiplets, whose index contribution is the same as $I_{N=1}$.}
\end{figure}
\begin{figure}[tbp]
%\vspace{0.3cm}
\centering % \begin{center}/\end{center} takes some additional vertical space
\includegraphics[width=.49\textwidth]{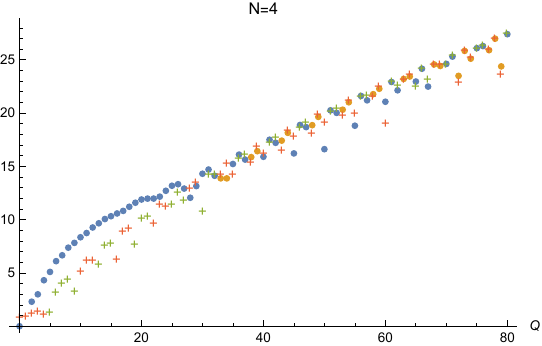}
\hfill
\includegraphics[width=.49\textwidth]{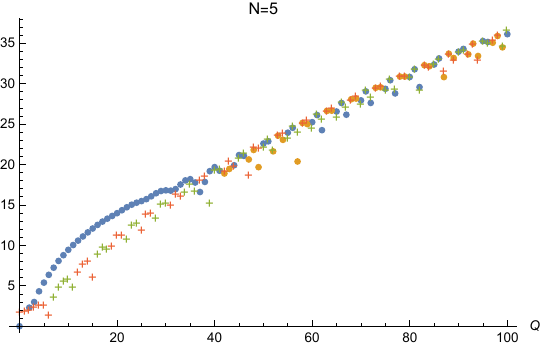}
\caption{\label{fig:n=45-im} $\log I_{N=4}/I_{N=1}(Q)$ (blue \& orange dots) vs $\tilde S(Q;\gamma = 1.03, \, \delta = 1.1)$ (green \& red crosses) on the left and $\log I_{N=5}/I_{N=1}(Q)$ (blue \& orange dots) vs $\tilde S(Q;\gamma = 1.24, \, \delta = 1.9)$ (green \& red crosses) on the right, both up to $x^{100}$. Blue dots and green crosses correspond to coefficients with the positive sign, whereas orange dots and red crosses correspond to coefficients with the negative sign. Other than the points for small $Q$ and those corresponding to $\cos[\dots] \approx 0$ in \eqref{eq:Stilde}, whose logarithm diverges, there is good agreement between the dots and the crosses.}
\end{figure}
Based on this idea, we aim to evaluate the indices for $N = 4, \, 5$ using truncated vortex contributions, as the exact results are more difficult to obtain due to the high computational cost. We calculate the indices keeping the vortex contributions up to vorticity 67 and
50 for $N = 4, \, 5$, respectively, which should be more than enough to capture the overall scaling of the
indices, according to the comparison made for $N = 2, \, 3$. The results are shown in Figures~\ref{fig:n=45}~and~\ref{fig:n=45-im}, where the latter includes the correction by the cosine factor in \eqref{eq:Stilde} as well.
As expected, we observe good agreement between the large-$N$ curves
and the approximated indices computed using truncated vortex contributions. However, it remains unclear whether this level of precision is sufficient for analyzing the complex-$\beta$ phase diagram for $N = 4, \, 5$, and we thus leave this investigation for future work.
\\

%\newpage
\section{BPS indices on Higgs vacua of the 3d ADHM with a single flavor}
\label{sec:Higgs}

As explained in the previous section, the superconformal index can be written as a sum of indices over the discrete Higgs vacua of the theory with generic real masses. Importantly, this structure is not unique to the superconformal index but also appears in other types of supersymmetric partition functions on manifolds that are $S^1$ fibered over $S^2$ \cite{Pasquetti:2011fj,Beem:2012mb,Hwang:2012jh,Taki:2013opa,Fujitsuka:2013fga,Benini:2013yva,Benini:2015noa,Choi:2019dfu,Colombo:2024mts}. A notable example is the topologically twisted index \cite{Benini:2015noa}, which exhibits the same structure. The building blocks in both cases are the same: the vortex partition function on $\mathbb R^2 \times S^1$ (including perturbative contribution). The only difference lies in how these building blocks are glued together. As we have seen, for the superconformal index, we glue two vortex partition functions related by $f \leftrightarrow f^{-1}$ where $f$ collectively denotes all the symmetry fugacities, while $m$, collectively denoting symmetry fluxes, remains invariant. In contrast, for the topologically twisted index, we glue two vortex partition functions related by $m \leftrightarrow -m$, whereas $f$ remains invariant \cite{Choi:2019dfu}.

In fact, the topologically twisted index admits an alternative formulation, called Bethe ansatz formalism, which expresses the index as a sum over solutions to the so-called Bethe ansatz equation (BAE), the equation of motion associated with the twisted superpotential of the theory compactified on $S^1$ \cite{Benini:2015noa}. For the ADHM quiver, such BAE solutions coincide precisely with the discrete Higgs vacua discussed earlier. The key point is that there is evidence suggesting that each BAE solution, or a certain set of solutions, corresponds to a particular holographic dual gravity solution. For instance, for the 4d $\mathcal N=4$ SYM index with equal angular velocities, there exist known BAE solutions, called the Hong-Liu solutions, whose contribution in the large-$N$ limit match the standard Euclidean black hole solutions and their orbifolds \cite{Aharony:2021zkr}. However, this correspondence is subtle, as it no longer holds when the angular velocities are unequal \cite{Aharony:2024ntg}.

Since we are discussing the unrefined index, it is natural to ask whether each Higgs vacuum, viewed as a version of a BAE ansatz solution, has any gravitational interpretation in the large-$N$ limit. Unfortunately, this is difficult to answer, as we currently lack the tools to evaluate the large-$N$ limit of the index for an individual Higgs vacuum. Instead, we aim to analyze the complex-$\beta$ phase diagram of the index on each Higgs vacuum to see if there is any relation between the peaks in the phase diagrams and the Higgs vacua of theory.

One immediate difficulty is that the index at an individual Higgs vacuum does not have well-defined unrefined limit. For example, the $U(3)$ ADHM with a single flavor has three Higgs vacua, for which the index reads as follows:
\ytableausetup{smalltableaux}
\begin{align}
\begin{ytableau}[] \\ \\ \\ \end{ytableau}&: \quad -q^\frac12 \, \frac{t^\frac12 z^4}{z-z^{-1}}+q \, \frac{t^{-1}z-z^4 \left(w+w^{-1}\right)-t z^5 \left(z^2+1+z^{-2}\right)}{z-z^{-1}}+\dots \,, \label{eq:linear 1} \\
\begin{ytableau}[] & \\ \\ \end{ytableau}&: \quad 1+q^\frac12 \left[t^{-\frac12} \left(w+w^{-1}\right)+t^\frac12 \left(z^3+2 z+2 z^{-1}+z^{-3}\right)\right] \nonumber \\
&\quad\quad +q \left[t^{-1} \left(2 w^2+1+2 w^{-2}\right)+\left(w+w^{-1}\right) \left(z^3+3 z+3 z^{-1}+z^{-3}\right)\right. \nonumber \\
&\qquad\quad\quad \left.+t \left(z^6+2 z^4+5 z^2+5+5 z^{-2}+2 z^{-4}+z^{-6}\right)\right]+\dots \,, \\
\begin{ytableau}[] & & \\ \end{ytableau}&: \quad q^\frac12 \, \frac{t^\frac12 z^{-4}}{z-z^{-1}}-q \, \frac{t^{-1} z^{-1}-z^{-4} \left(w+w^{-1}\right)-t z^{-5} \left(z^2+1+z^{-2}\right)}{z-z^{-1}}+\dots \,, \label{eq:linear 2}
\end{align}
whose sum results in the complete superconformal index:
\begin{align}
&\mathcal I_{N=3}(w,z,t,q) = 1+ q^\frac12 \left[t^{-\frac12} \left(w+w^{-1}\right)+t^\frac12 \left(z+z^{-1}\right)\right] \nonumber \\
&\quad+q \left[2 t^{-1} \left(w^2+1+w^{-2}\right)+2 \left(w+w^{-1}\right) \left(z+z^{-1}\right)+2 t \left(z^2+1+z^{-2}\right)\right]+\dots \,.
\end{align}
One can see that the indices for $\tiny\begin{ytableau}[] & & \\ \end{ytableau}$ and $\tiny\begin{ytableau}[] & & \\ \end{ytableau}^T$, which are symmetric under $z \leftrightarrow z^{-1}$, include a factor of $(z-z^{-1})^{-1}$, which diverges in the limit $z \rightarrow 1$. The origin of this factor is zero modes of an adjoint field at the corresponding Higgs vacua, whose contribution should diverge in the massless limit. This is not surprising because we have \emph{discrete} Higgs vacua only when real masses are nonzero, while they become continuous vacuum moduli space in the massless limit.

Obviously, such zero modes are not part of the BPS spectrum in the massless limit
leading to the SCFT. Namely, they must be paired up to form a long multiplet, whose combined
contribution to the index vanishes. We find that certain combinations of the Higgs vacuum
indices yield those with integer coefficients, where all the divergent contributions from the zero modes
cancel out. Here are some examples for low values of $N$:
\begin{align}
&N=2: \qquad \begin{ytableau}[] \\ \\ \end{ytableau}+\begin{ytableau}[] & \\ \end{ytableau} \,, \\
&N=3: \qquad \begin{ytableau}[] \\ \\ \\ \end{ytableau}+\begin{ytableau}[] & & \\ \end{ytableau} \,, \qquad \begin{ytableau}[] & \\ \\ \end{ytableau} \\
&N=4: \qquad \begin{ytableau}[] \\ \\ \\ \\ \end{ytableau}+\begin{ytableau}[] & & & \\ \end{ytableau} \,, \qquad \begin{ytableau}[] & \\ \\ \\ \end{ytableau}+\begin{ytableau}[] & \\ & \\ \end{ytableau}+\begin{ytableau}[] & & \\ \\ \end{ytableau} \,, \\
&N=5: \qquad \begin{ytableau}[] \\ \\ \\ \\ \\ \end{ytableau}+\begin{ytableau}[] & & & & \\ \end{ytableau} \,, \qquad \begin{ytableau}[] & \\ \\ \\ \\ \end{ytableau}+\begin{ytableau}[] & & & \\ \\ \end{ytableau} \,, \qquad \begin{ytableau}[] & \\ & \\ \\ \end{ytableau}+\begin{ytableau}[] & & \\ \\ \\ \end{ytableau}+\begin{ytableau}[] & & \\ & \\ \end{ytableau} \,.
\end{align}
Note that all the combinations are symmetric under $z \leftrightarrow z^{-1}$, which is a symmetry of the SCFT. In fact, for $N = 4, \, 5$, the two linear Young diagrams corresponding to the partitions $(N)$ and $(1^N)$ are already free of zero-mode contributions individually, at least at low orders in $q$-expansion. Nevertheless, we sum their contributions to ensure symmetry under $z \leftrightarrow z^{-1}$ for the refined index, as we do for the other combinations.
We will call such indices without zero-modes contributions \emph{smooth} Higgs vacuum indices, in the sense that they admit a smooth unrefined limit.

One should note that the smooth indices also include states that become
non-BPS in the SCFT limit. Nevertheless, it is possible that such accidental BPS
states in the gapped vacua are subdominant, and each smooth Higgs vacuum index
could still provide insight into the primary contribution to the SCFT index, or
holographically, that into a possible gravity solution in connection with the BAE solutions.

\begin{figure}[tbp]
\centering % \begin{center}/\end{center} takes some additional vertical space
\includegraphics[width=.6\textwidth]{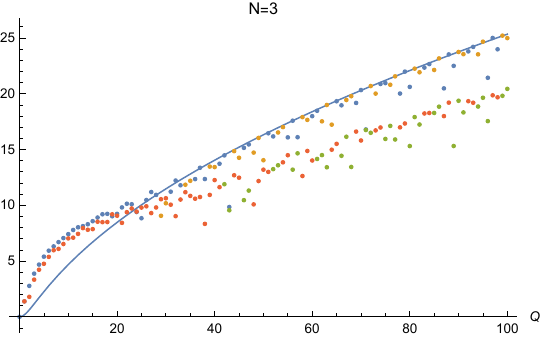}
\caption{\label{fig:n=3-Hig} Two smooth Higgs vacuum indices for $N = 3$. The index corresponding to $\tiny\begin{ytableau}[] & \\ \\ \end{ytableau}$ are represented by blue (positive sign) and orange (negative sign) dots, whereas the index corresponding to $\tiny\begin{ytableau}[] & & \\ \end{ytableau}+\begin{ytableau}[] & & \\ \end{ytableau}^T$ are represented by green (positive sign) and red (negative sign) dots. While both of them exhibit oscillating behavior, only the index for $\tiny\begin{ytableau}[] & \\ \\ \end{ytableau}$ matches the large-$N$ entropy $\tilde S$ (blue line).}
\end{figure}
In this sense, it is natural to ask whether the peaks in the complex-$\beta$ phase diagram we observed in the previous section are related to the smooth Higgs vacuum indices. The answer seems to be yes. We focus on the $N = 3$ case since, for $N = 2$, there is only one smooth Higgs vacuum index for $N = 2$, which is nothing but the superconformal index, and for $N = 4, \, 5$, we only have approximate indices. For $N = 3$, there are two smooth Higgs vacuum indices: one corresponding to the Young diagram $\tiny\begin{ytableau}[] & \\ \\ \end{ytableau}$ and the other given by the sum of the Higgs vacuum indices associated with $\tiny\begin{ytableau}[] & & \\\end{ytableau}$ and $\tiny\begin{ytableau}[] & & \\ \end{ytableau}^T$. The growth of each smooth Higgs vacuum index is shown in Figure~\ref{fig:n=3-Hig}.

\begin{figure}[tbp]
\centering % \begin{center}/\end{center} takes some additional vertical space
\includegraphics[width=.31\textwidth]{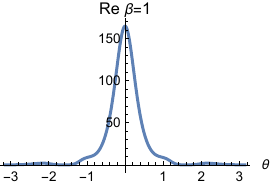}
\hfill
\includegraphics[width=.31\textwidth]{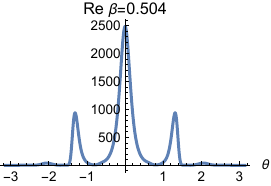}
\hfill
\includegraphics[width=.31\textwidth]{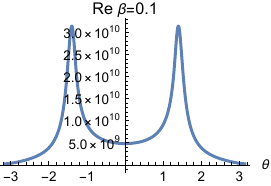}
\caption{\label{fig:n=3-Hig1} The complex-$\beta$ phase diagram of the smooth Higgs vacuum index for $\tiny\begin{ytableau}[] & \\ \\ \end{ytableau}$, evaluated at $\mathrm{Re} \, \beta = 1, \, 0.504, \, 0.1$. The minor peak around $\theta = \pm2.04$ observed in the phase diagram of the complete superconformal index (Figure~\ref{fig:n=3-phase}) does not appear here.}
\vspace{0.4cm}
\centering % \begin{center}/\end{center} takes some additional vertical space
\includegraphics[width=.31\textwidth]{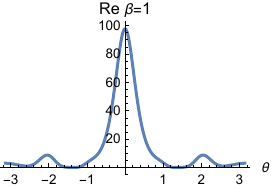}
\hfill
\includegraphics[width=.31\textwidth]{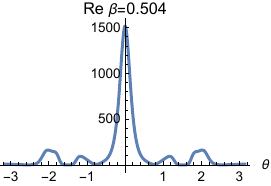}
\hfill
\includegraphics[width=.31\textwidth]{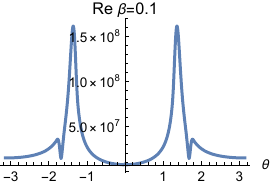}
\caption{\label{fig:n=3-Hig2} The complex-$\beta$ phase diagram of the smooth Higgs vacuum index for $\tiny\begin{ytableau}[] & & \\\end{ytableau}+\tiny\begin{ytableau}[] & & \\ \end{ytableau}^T$, evaluated at $\mathrm{Re} \, \beta = 1, \, 0.504, \, 0.1$. One can observe the same small peak around $\theta = \pm2.04$ that appears in the phase diagram of the complete superconformal index (Figure~\ref{fig:n=3-phase}). However, for small $\mathrm{Re} \, \beta$, eventually it is dominated by the peak around $\theta = \pm1.40$ corresponding to the known black hole contribution.}
\end{figure}
Both smooth indices exhibit a transition from a non-oscillating region to an oscillating region. However, the slope in the oscillating region differs for each Higgs vacua. Only the index for $\tiny\begin{ytableau}[] & \\ \\ \end{ytableau}$ matches the large-$N$ entropy $\tilde S$, which clearly distinguishes the two indices. We therefore analyze the complex-$\beta$ phase diagrams of the two smooth Higgs vacuum indices individually to determine whether the phase diagrams also display different patterns for different Higgs vacua. The results are shown in Figures~\ref{fig:n=3-Hig1}~and~\ref{fig:n=3-Hig2}, respectively.

Most interestingly, the small peak observed in the phase diagram of the superconformal index appears only in that of the smooth index associated with $\tiny\begin{ytableau}[] & & \\ \end{ytableau}+\begin{ytableau}[] & & \\ \end{ytableau}^T$. The phase diagram for $\tiny\begin{ytableau}[] & \\ \\ \end{ytableau}$ shows only a tiny bump, which is negligible compared to the other peaks. Although taking the unrefined limit causes us to lose detailed charge information about the states, the Young diagrams still hint at which charge sectors dominate at each Higgs vacuum. For instance, comparing the Higgs vacuum indices in \eqref{eq:linear 1} and \eqref{eq:linear 2} associated with the linear Young diagrams, we find that \eqref{eq:linear 1} tends to have positive powers of $z$, corresponding to the excitation of the adjoint field $Y$, while \eqref{eq:linear 2} tends to have negative powers, corresponding to $Z$. It is thus natural to expect that a linear Young diagram captures states with biased charges, either $Q_1 \gg Q_2$ or $Q_1 \ll Q_2$ where $Q_1$ and $Q_2$, two of the Cartans of $Spin(8)_R$, count roughly the excitations of the adjoint chirals $Y$ and $Z$, respectively. This is in contrast to the standard black hole peak, which is expected to be dominated by states with approximately equal charges $Q_1 \approx Q_2$, mainly arising from $\tiny\begin{ytableau}[] & \\ \\ \end{ytableau}$.\footnote{This is consistent with the conjecture from the topologically twisted index that the black hole saddle arises from the triangular-shaped Young diagram \cite{Crew:2024fom}.} Therefore, we need to investigate the biased charge sector to interpret the contribution of the small peak arising from Higgs vacua corresponding to the linear Young diagrams. A precise statement, however, requires the refined index with the complete charge information, which is beyond our current computational capabilities. It would be interesting to extend the refined computation to higher orders to better understand the nature of this small peak. In the next section, we will provide further evidence of this expectation using particular limits of the superconformal index, leading to the Hilbert series and the half-BPS indices of the ADHM quiver.

Another crucial observation is that even for the linear Young diagrams, the standard black hole peak dominates when $\mathrm{Re} \, \beta$ is small. See the right plot in Figure~\ref{fig:n=3-Hig2}. Thus, we expect that although the slope of the index for $\tiny\begin{ytableau}[] & & \\ \end{ytableau}+\begin{ytableau}[] & & \\ \end{ytableau}^T$ differs from that of the black hole entropy in the charge range up to $Q = 100$, the contribution from the standard black hole solution should dominate at larger $Q$. It would be interesting to extend the computation to higher orders to confirm this expectation. In addition, this suggests that the contribution of the small peak cannot be seen by merely tracing the leading term in the large-$N$ limit, which will be dominated by the standard black hole solutions. This is, in fact, consistent with the Cardy-limit computation in \cite{Choi:2019zpz}, where it was shown that, in the Cardy limit, the superconformal index reduces to the contribution of the linear Young diagrams. Therefore, to extract the information of the small peak, one should look at a restricted charge sector where the standard black hole contribution is suppressed.
\\

\begin{figure}[tbp]
\centering % \begin{center}/\end{center} takes some additional vertical space
\includegraphics[width=.49\textwidth]{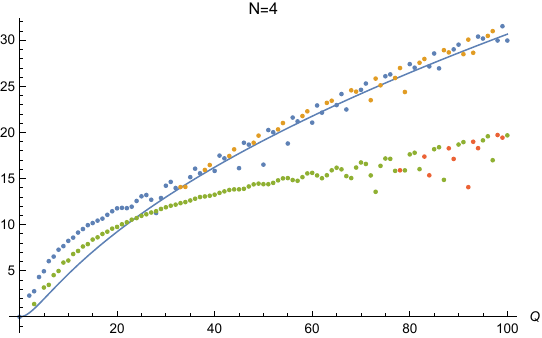}
\hfill
\includegraphics[width=.49\textwidth]{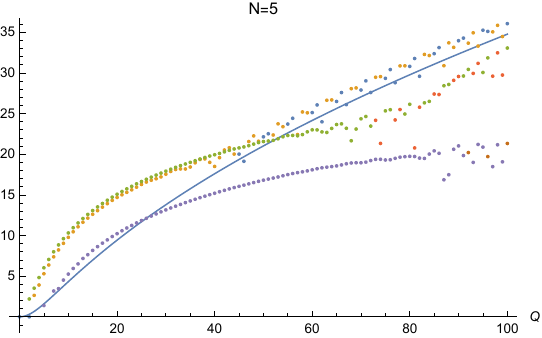}
\caption{\label{fig:n=4-Hig} Two smooth Higgs vacuum indices for $N = 4$ with vortex contributions up to vorticity 67 (left) and three for $N = 5$ with vortex contributions up to vorticity 50 (right). For $N = 4$, the index corresponding to $\tiny\begin{ytableau}[] & \\ & \\ \end{ytableau}+\tiny\begin{ytableau}[] & & \\ \\ \end{ytableau}+\tiny\begin{ytableau}[] & & \\ \\ \end{ytableau}^T$ are represented by blue (positive sign) and orange (negative sign) dots, whereas the index corresponding to $\tiny\begin{ytableau}[] & & & \\ \end{ytableau}+\begin{ytableau}[] & & & \\ \end{ytableau}^T$ are represented by green (positive sign) and red (negative sign) dots. For $N = 5$, the index corresponding to $\tiny\begin{ytableau}[] & & \\ \\ \\ \end{ytableau}+\tiny\begin{ytableau}[] & & \\ & \\ \end{ytableau}+\tiny\begin{ytableau}[] & & \\ & \\ \end{ytableau}^T$ are represented by blue (positive sign) and orange (negative sign) dots; the index corresponding to $\tiny\begin{ytableau}[] & & & \\ \\ \end{ytableau}+\begin{ytableau}[] & & & \\ \\ \end{ytableau}^T$ by green (positive sign) and red (negative sign) dots; and the index corresponding to $\tiny\begin{ytableau}[] & & & & \\ \end{ytableau}+\begin{ytableau}[] & & & & \\ \end{ytableau}^T$ by purple (positive sign) and brown (negative sign) dots. In both cases, the large-$N$ entropy $\tilde S$ is shown by the blue lines.}
\end{figure}
The indices for $N = 4, \, 5$ are, on the other hand, calculated using the truncated vortex contributions, which lose the fine details of the index. For this reason, we relegate the analysis of their complex-$\beta$ phase diagrams to future study and instead provide only comparisons of the overall growth of smooth Higgs vacuum indices, which show similar patterns to the $N=3$ case. See Figure~\ref{fig:n=4-Hig}.
\\

\section{Factorization of the 3d ADHM index with multiple flavors}
\label{sec:factorization}

While we have so far discussed the ADHM quiver with a single flavor, one can also consider the theory with $F$ flavors. This theory arises as a UV gauge theory of the 3d SCFT describing the low-energy dynamics of M2-branes probing the singularity $\mathbb C^2 \times \mathbb C^2/\mathbb Z_F$ \cite{Porrati:1996xi}, whose holographic dual is M-theory on AdS$_4 \times S^7/\mathbb Z_{F}$. Famously, the Higgs branch of the ADHM quiver is the moduli space of $N$ instantons for the $U(F)$ gauge theory \cite{Atiyah:1978ri}, while its Coulomb branch is given by the symmetric product of $N$ copies of the orbifold $\mathbb C^2/\mathbb Z_F$, i.e., $\mathrm{Sym}^N (\mathbb C^2/\mathbb Z_F)$  \cite{deBoer:1996mp}, which is the same as the $N$-instanton moduli space when $F = 1$. The Higgs and Coulomb branches are no longer identical when $F > 1$, indicating that the theory is not self-mirror, unlike the single-flavor case. Instead, its mirror dual is a circular quiver of $F$ $U(N)$ gauge nodes with one flavor attached to a single node, which is a special case of the Kronheimer--Nakajima quiver \cite{PeterBKronheimer:1990zmj}. In addition, as evident from the geometry, the supersymmetry is not enhanced and remains as $\mathcal N=4$ in the IR if $F > 1$.

One can write down the index for the theory with multiple flavors, again using the matrix integral formula:
\begin{align}
\label{eq:matrix_multi}
& \mathcal I(w,z,t,q) = \\
& \frac{1}{N!} \sum_{m \in \mathbb Z^N/S_N} \oint \left(\prod_{a = 1}^N 
\frac{d s_a}{2 \pi i s_a} w^{m_a} t^{-F |m_a|/2} q^{F |m_a|/2}\right) \times \nonumber \\
&\left(\prod_{1 \leq a \neq b \leq N} \left(1-s_a s_b^{-1} q^{|m_a-m_b|}\right)\right) \left(\prod_{i = 1}^F \prod_{a = 1}^N \frac{(s_a^{-1} y_i t^{-\frac{1}{2}} q^{\frac{3}{2}+|m_a|};q^2)}{(s_a y_i^{-1} t^\frac{1}{2} q^{\frac{1}{2}+|m_a|};q^2)} \frac{(s_a y_i^{-1} t^{-\frac{1}{2}} q^{\frac{3}{2}+|m_a|};q^2)}{(s_a^{-1} y_i t^\frac{1}{2} q^{\frac{1}{2}+|m_a|};q^2)}\right) \times \nonumber \\
&\left(\prod_{a,b = 1}^N \frac{(s_a^{-1} s_b t q^{1+|-m_a+m_b|};q^2) (s_a^{-1} s_b z^{-1} t^{-\frac{1}{2}} q^{\frac{3}{2}+|-m_a+m_b|};q^2) (s_a^{-1} s_b z t^{-\frac{1}{2}} q^{\frac{3}{2}+|-m_a+m_b|};q^2)}{(s_a s_b^{-1} t^{-1} q^{1+|m_a-m_b|};q^2) (s_a s_b^{-1} z t^\frac{1}{2} q^{\frac{1}{2}+|m_a-m_b|};q^2) (s_a s_b^{-1} z^{-1} t^\frac{1}{2} q^{\frac{1}{2}+|m_a-m_b|};q^2)}\right)
\end{align}
where we have again used the shorthand $(a;q^2)$ for the q-Pochammer symbol $(a;q^2)_\infty = \prod_{k = 0}^\infty (1-a q^{2 k})$. On top of the symmetries of the single-flavor case shown in Table~\ref{tab:sym}, now we also have the $SU(F)$ flavor symmetry under which the flavors form the fundamental representation, whose fugacities $y_i$ satisfy $\prod_{i = 1}^F y_i = 1$. 

This integral formula is useful to study the large-$N$ limit of the index. For instance, the large-$N$ free energy can be obtained using the Cardy block formalism \cite{Choi:2019dfu}:
\begin{align}
\label{eq:largeN_multi}
\mathcal F \; = \;  -\log \mathcal I \; \approx \; i \frac{4 \sqrt2 F^\frac12 N^\frac32}{3} \frac{\sqrt{\Delta_1 \Delta_2 \Delta_3 \Delta_4}}{2 \beta} \,,
\end{align}
where we have an additional factor of $F^\frac12$ compared to that of the single-flavor theory.
Note that there is no $y_i$-dependence in this leading contribution. Despite its usefulness in the large-$N$ analysis, however, the matrix integral formula is inefficient for computing finite-$N$ indices. Thus, our goal here is to derive an alternative formula using the factorization of the superconformal index for multiple flavors.
\\

The derivation is almost identical to that of the single-flavor case; therefore, we explain it focusing on the differences. In \eqref{eq:matrix_multi}, the integration contour for each $s_a$ is taken as the unit circle. Then we can evaluate this integral by taking the residues at the poles outside the contour. Assuming $|q| < 1$ and $|t| = |z| = |y_i| = 1$, the poles sitting outside the unit circle are determined as the intersections of the following hyperplanes:
\begin{align}
s_a &= y_i t^{-\frac{1}{2}} q^{-\frac{1}{2}} q^{-|m_a|-2 k_a}, \\
s_a &= s_b z^{-1} t^{-\frac{1}{2}} q^{-\frac{1}{2}} q^{-|m_a-m_b|-2 k_a}, \\
s_a &= s_b z t^{-\frac{1}{2}} q^{-\frac{1}{2}} q^{-|m_a-m_b|-2 k_a}, \\
s_a &= s_b t q^{-1} q^{-|m_a-m_b|-2 k_a}
\end{align}
with $k_a \geq 0$. As noted in \cite{Choi:2019zpz}, a pole intersecting the last type of hyperplanes has the vanishing residue, and the relevant poles are only those determined by the other three. Thus, the residue calculation is essentially the same as that of two-adjoint theories, whose poles are generically labeled by binary trees \cite{Hwang:2018uyj}. The root node represents the first-type hyperplane, and its two child nodes, as well as their descendants, correspond to the second and the third-type hyperplanes.

In \cite{Choi:2019zpz}, it was argued that for the 3d ADHM theory with a single flavor, the poles having nontrivial residues are only those whose corresponding binary tree can be mapped to a Young diagram. Under this mapping, the initial box of the Young diagram, located at position $(1,1)$, represents the root node of the binary tree, while each box that is vertically or horizontally adjacent to a previously assigned box corresponds to a child node of that node, defined recursively. This reflects the facts that the trees having no counterpart Young diagrams have vanishing residues, and that the would-be higher order poles are rendered simple by additional zeros arising from the superpotential constraints on the fugacities. The important point is that this argument only relates to how the second and the third-type hyperplanes intersect, and it is independent of the first-type hyperplane. Thus, as long as each $y_i$ has a generic value so that no accidental pole or zero arises, it is still valid for multiple flavors, as the number of flavors only affects the first-type hyperplanes.

More precisely, for $F$ flavors, we have $F$ hyperplanes of the first type distinguished by $y_i$. Thus, the poles are now labeled by a set of $F$ colored Young diagrams of $N$ boxes in total. Other than this, the remaining calculation is the same as that of the single-flavor case, which in the end results in the following form of the index:
\begin{align}
\label{eq:fact_index_multi}
\mathcal I(\vec y,w,z,t,q) = \sum_{\sum_{i =1}^F |\mathcal Y_i| = N} Z_\text{pert}^{\vec{\mathcal Y}}(\vec y,z,t,q) \, Z_\text{vort}^{\vec{\mathcal Y}}(\vec y,w,z,t,q) \, Z_\text{vort}^{\vec{\mathcal Y}} (\vec y^{-1},w^{-1},z^{-1},t^{-1},q^{-1})
\end{align}
with
\begin{align}
Z^{\vec{\mathcal Y}}_\text{pert}(\vec y,z,t,q) &= \left(\prod_{\mathsf a\neq \mathsf b \in \vec{\mathcal Y}} \left(1-v_\mathsf a^{-1} v_{\mathsf b}\right)\right) \left(\prod_{i = 1}^F \prod_{\mathsf a \in \vec{\mathcal Y}} \frac{(v_\mathsf a y_i q^2;q^2)}{(v_\mathsf a^{-1} y_i^{-1};q^2)} \frac{(v_\mathsf a^{-1} y_i^{-1} (t q)^{-1} q^2;q^2)}{(v_\mathsf a y_i t q;q^2)}\right) \nonumber \\
&\quad \times \left(\prod_{\mathsf a,\mathsf b \in \vec{\mathcal Y}} \frac{(v_\mathsf a v_\mathsf b^{-1} t q;q^2) (v_\mathsf a v_\mathsf b^{-1} (z t^\frac12 q^\frac12)^{-1} q^2;q^2) (v_\mathsf a v_\mathsf b^{-1} (z^{-1} t^\frac12 q^\frac12)^{-1} q^2;q^2)}{(v_\mathsf a^{-1} v_\mathsf b (t q)^{-1} q^2;q^2) (v_\mathsf a^{-1} v_\mathsf b z t^\frac12 q^\frac12;q^2) (v_\mathsf a^{-1} v_\mathsf b z^{-1} t^\frac12 q^\frac12;q^2)}\right) ,
\end{align}
\begin{align}
Z_\text{vort}^{\vec{\mathcal Y}}(\vec y,w,z,t,q) &= \sum_{k_{\mathsf a}}
\left(w t^\frac{F}{2} q^{-\frac{F}{2}}\right)^{\sum_{\mathsf a \in \vec{\mathcal Y}} k_{\mathsf a}} \left(\prod_{i = 1}^F \prod_{\mathsf a \in \vec{\mathcal Y}} \frac{(v_{\mathsf a}^{-1} y_i^{-1};q^2)_{-k_{\mathsf a}}}{(v_{\mathsf a}^{-1} y_i^{-1} t^{-1} q;q^2)_{-k_{\mathsf a}}}\right) \nonumber \\
&\quad \times \left(\prod_{\mathsf a \neq \mathsf b \in \vec{\mathcal Y}} \frac{(v_{\mathsf a}^{-1} v_{\mathsf b} z t^\frac12 q^\frac12;q^2)_{-k_{\mathsf a}+k_{\mathsf b}} (v_{\mathsf a}^{-1} v_{\mathsf b} t^{-1} q;q^2)_{-k_{\mathsf a}+k_{\mathsf b}} }{(v_{\mathsf a}^{-1} v_{\mathsf b};q^2)_{-k_{\mathsf a}+k_{\mathsf b}} (v_{\mathsf a}^{-1} v_{\mathsf b} z t^{-\frac12} q^\frac32;q^2)_{-k_{\mathsf a}+k_{\mathsf b}}}\right)
\end{align}
where the vanishing factors appearing in the expression must be discarded.
Here, $v_\mathsf a$ for $\mathsf a \in \mathcal Y_i$ is given by
\begin{align}
v_\mathsf a = y_i^{-1} z^{\mathsf i (\mathsf a)-\mathsf j (\mathsf a)} (t q)^{\frac{1}{2} (\mathsf i (\mathsf a)+\mathsf j (\mathsf a)-2)}
\end{align}
where $(\mathsf i (\mathsf a),\mathsf j (\mathsf a))$ is the position of box $\mathsf a$ in Young diagram $\mathcal Y_i$. $k_\mathsf a$, a combination of $\vec m$ and $\vec k$, is a non-negative integer assigned to each $\mathsf a$ such that those integers are non-decreasing along each row and column of $\mathcal Y_i$ starting from the initial box. See \cite{Choi:2019zpz} for further details of the computations.

In the remainder of this section, we will give a brief discussion of examples and also examine a useful limit of this expression, which leads to a simple formula for the refined Higgs branch Hilbert series, capturing BPS states annihilated by two complex supercharges.

\subsection{Examples}

As an example, let us consider the $U(2)$ ADHM with two flavors. With generic real masses, the theory has five discrete Higgs vacua labeled by the following colored Young diagrams:
\begin{align}
\label{eq:Young22}
\begin{ytableau}[]
*(blue!67) \\
*(blue!67)
\end{ytableau}
\quad\oplus\quad
\begin{ytableau}[]
*(blue!67) & *(blue!67)
\end{ytableau}
\quad\oplus\quad
\begin{ytableau}[]
*(red!67) \\
*(red!67)
\end{ytableau}
\quad\oplus\quad
\begin{ytableau}[]
*(red!67) & *(red!67)
\end{ytableau}
\quad\oplus\quad
\begin{ytableau}[]
*(blue!67)
\end{ytableau}
\quad
\begin{ytableau}[]
*(red!67)
\end{ytableau}
\end{align}
where blue Young diagrams are associated with $y_1 = y$, while red Young diagrams correspond to $y_2 = 1/y$. The index evaluated at each vacuum is given by
\begin{align}
\begin{ytableau}[]
*(blue!67) \\
*(blue!67)
\end{ytableau}
&: \quad q^\frac12 \, \frac{y^{-3} z^2}{\left(y-y^{-1}\right) \left(z-z^{-1}\right)}+q \left[ -\frac{y^{-2} z}{z-z^{-1}}+\frac{y^{-4} z^3 \left(y+y^{-1}\right)}{\left(y-y^{-1}\right) \left(z-z^{-1}\right)} \right] \nonumber \\
&\quad\quad  +q^{3/2} \left[ -\frac{y^{-3} z^2 (y+y^{-1})}{z-z^{-1}}+\frac{y^{-7} z^4+y^{-4} z^3 \left(y+y^{-1}\right) \left(z+z^{-1}\right)}{\left(y-y^{-1}\right) \left(z-z^{-1}\right)}\right. \nonumber \\
&\qquad\qquad\quad\quad  \left. +\frac{y^{-2} z^2 \left(z+z^{-1}\right)}{2} \left( \frac{y+y^{-1}}{y-y^{-1}}-\frac{z+z^{-1}}{z-z^{-1}}\right)\right] +\dots \\
\begin{ytableau}[]
*(blue!67) & *(blue!67)
\end{ytableau}
&: \quad \left(z \leftrightarrow z^{-1}\right) \\
\begin{ytableau}[]
*(red!67) \\
*(red!67)
\end{ytableau}
&: \quad \left(y \leftrightarrow y^{-1}\right) \\
\begin{ytableau}[]
*(red!67) & *(red!67)
\end{ytableau}
&: \quad \left(y \leftrightarrow y^{-1} \; \& \; z \leftrightarrow z^{-1}\right) \\
\begin{ytableau}[]
*(blue!67)
\end{ytableau}
\quad
\begin{ytableau}[]
*(red!67)
\end{ytableau}
&: \quad 1+q^\frac12 \left(1+\chi_3\right) \left(z+z^{-1}\right)+q \left[2+2 \chi_3+\left(4+\chi_3+\chi_5\right) \left(z^2 + 1 + z^{-2}\right)\right] \nonumber \\
&\quad\quad +q^\frac32 \left[\left(4+4 \chi_3+2 \chi_5\right) \left(z+z^{-1}\right)+\left(2+2 \chi_3+\chi_5+\chi_7\right) \left(z^3+z+z^{-1}+z^{-3}\right)\right] \nonumber \\
&\quad\quad +\dots
\end{align}
where one can use the permutation symmetries $y \leftrightarrow y^{-1}$ and $z \leftrightarrow z^{-1}$ to obtain other three indices from the first one. In the last line, we have used the $SU(2)$ flavor character $\chi_n = \sum_{k=-\frac{n-1}{2}}^{\frac{n-1}{2}} y^{2 k}$ of dimension $n$.
By summing these contributions, one obtains the complete superconformal index as follows:
\begin{align}
I_{N=2;F=2}(\vec y,w,z,t,q) &= 1+q^\frac12 \left(z+z^{-1}\right)+q \left(3+\chi_3+2 \left(z^2+1+z^{-2}\right)\right) \nonumber \\
&\quad +q^\frac32 \left(\left(6+2 \chi_3\right) \left(z+z^{-1}\right)+2 \left(z^3+z+z^{-1}+z^{-3}\right)\right)+\dots \,.
\end{align}
The indices for other values of $N$ and $F$ can also be obtained using the formula \eqref{eq:fact_index_multi}.

\begin{figure}[tbp]
\centering % \begin{center}/\end{center} takes some additional vertical space
\includegraphics[width=.6\textwidth]{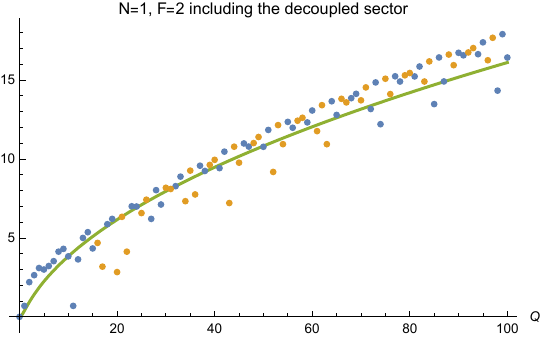} \\
\vspace{0.6cm}
\includegraphics[width=.6\textwidth]{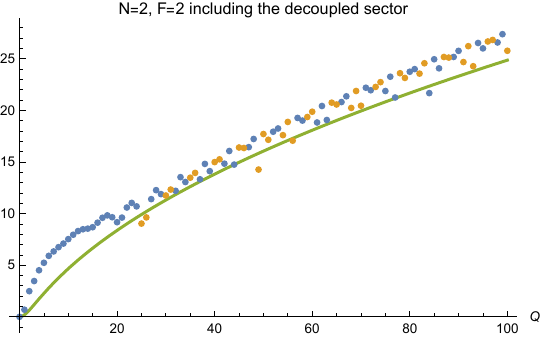} \\
\vspace{0.6cm}
\includegraphics[width=.6\textwidth]{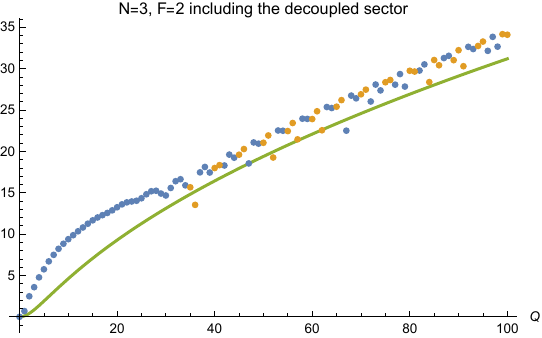}
\caption{\label{fig:f=2} $\log I_{N=1, \, 2, \, 3; \, F=2}(Q)$ (blue \& orange dots) vs $S_{N=1,\, 2,\, 3; \, F=2}(Q)$ (green line) up to $x^{100}$. Blue and orange dots represent coefficients with positive and negative signs, respectively. The agreement improves with increasing $N$.}
\end{figure}
One may also attempt to compare finite-$N$ indices of multi-flavor cases with the result from the large-$N$ analysis. Recall that the large-$N$ limit of the index with multiple flavors is governed by free energy \eqref{eq:largeN_multi}. Since the only difference is the additional factor of $F^\frac12$, one can easily obtain the large-$N$ entropy as follows:
\begin{align}
S_{N;F}(Q) = S(Q;F^\frac12 N^\frac32)
\end{align}
where $S$ on the right hand side is the entropy for a single flavor given in \eqref{eq:S}. Now we can compare $S_{N;F}$ with the logarithm of the unrefined indices at finite $N$ with multiple flavors, whose results are shown in Figure~\ref{fig:f=2} for $N = 1, \, 2, \, 3$ and $F = 2$.
Again, we observe that
the large-$N$ entropy provides a good approximation to the finite-$N$ indices. While a detailed examination of the complex-$\beta$ phase diagrams of these indices would also be of interest, we leave this to future
work.

\subsection{Limit to the Higgs branch Hilbert series}

The Hilbert series is the observable that encodes the data of (a branch of) the moduli space of vacua as an algebraic variety. For 3d $\mathcal N=4$ theories, one can mainly consider two types of it: the Higgs branch Hilbert series and the Coulomb branch Hilbert series. In particular, for `good' $\mathcal N=4$ gauge theories in the sense of Gaiotto--Witten, these Hilbert series can be obtained as particular limits of the superconformal index \cite{Razamat:2014pta}.\footnote{For 3d $\mathcal N=2$ theories, the distinction between the Higgs and Coulomb branches are not always clear. Nevertheless, one can obtain the Hilbert series of a certain `component' of the moduli space by taking a suitable limit of the superconformal index \cite{Hanany:2015via}.} In our case, the limits are given by
\begin{align}
\text{Higgs branch limit}&: \quad q, \, t^{-1} \; \rightarrow \; 0 \,, \quad  q t \text{ fixed,} \\
\text{Coulomb branch limit}&: \quad q, \, t \quad \; \rightarrow \; 0 \,, \quad  q/t \text{ fixed.}
\end{align}
Roughly speaking, the Higgs limit removes the contribution of monopole operators from the index while retaining that of the adjoint hypermultiplet, whereas the Coulomb limit removes the contribution of the adjoint hypermultiplet while retaining that of monopole operators. This is reflected to the BPS condition. Recall that the BPS condition for the states contributing the superconformal index satisfy
\begin{align}
E = \frac12 \left(Q_1+Q_2+Q_3+Q_4\right)+j \,,
\end{align}
where $Q_i$'s are integer-quantized Cartans of the $Spin(8)_R$ symmetry.
On the other hand, the states contributing to the Hilbert series are annihilated by two complex supercharges and satisfy the following relation:
\begin{align}
\text{Higgs branch limit}&: \quad E \; = \; \frac12 \left(Q_1+Q_2\right) , \qquad Q_3 = Q_4 = j = 0 \,, \\
\text{Coulomb branch limit}&: \quad E \; = \; \frac12 \left(Q_3+Q_4\right) , \qquad Q_1 = Q_2 = j = 0 \,.
\end{align}

Among these two limits, we focus on the Higgs branch limit, in which our factorized
index naturally leads to a simple formula for the Higgs branch Hilbert series. First, it is straightforward to show that, in the $w$-expansion of $Z_\text{vort}^{\vec{\mathcal Y}}$ in \eqref{eq:vort}, each power of $w$ is accompanied by at least a factor of $q/t$, which vanishes in the Higgs limit. Consequently, $Z_\text{vort}^{\vec{\mathcal Y}}$ goes to 1, and we are left only with $Z_\text{pert}^{\vec{\mathcal Y}}$. In the same limit, $Z_\text{pert}^{\vec{\mathcal Y}}$ becomes
\begin{align}
&\lim_{\substack{q, t^{-1} \rightarrow 0 \\ r \equiv (q t)^\frac12 \text{ fixed}}} Z_\text{pert}^{\vec{\mathcal Y}}(\vec y,z,q) \\
&\rightarrow \quad H^{\vec{\mathcal Y}}(\vec y,z,r) = \left(\prod_{\mathsf a\neq \mathsf b \in \vec{\mathcal Y}} \left(1-v_\mathsf a^{-1} v_\mathsf b\right)\right) \left(\prod_{i = 1}^F \prod_{\mathsf a \in \vec{\mathcal Y}} \frac{1}{(1-v_\mathsf a^{-1} y_i^{-1})} \frac{1}{(1-v_\mathsf a y_i r^2)}\right) \nonumber \\
&\qquad\qquad\qquad\qquad \times \left(\prod_{\mathsf a,\mathsf b \in \vec{\mathcal Y}} \frac{(1-v_\mathsf a v_\mathsf b^{-1} r^2)}{(1-v_\mathsf a^{-1} v_\mathsf b z r) (1-v_\mathsf a^{-1} v_\mathsf b z^{-1} r)}\right)
\end{align}
where the vanishing factors must be discarded as in the case of the superconformal index. $v_\mathsf a$ for $\mathsf a \in \mathcal Y_i$ is now given by 
\begin{align}
v_\mathsf a = y_i^{-1} z^{\mathsf i (\mathsf a)-\mathsf j (\mathsf a)} r^{\mathsf i (\mathsf a)+\mathsf j (\mathsf a)-2}
\end{align}
where, as before, $(\mathsf i (\mathsf a),\mathsf j (\mathsf a))$ is the position of the box $\mathsf a$ in Young diagram $\mathcal Y_i$. As a result, the (refined) Higgs branch Hilbert series is given by the following simple formula:
\begin{align}
\label{eq:HS}
\mathcal H_{N;F} (\vec y,z,r) = \sum_{\sum_{i =1}^F |\mathcal Y_i| = N} H^{\vec{\mathcal Y}}(\vec y,z,r) \,,
\end{align}
which generalizes the corresponding formula for $F=1$ derived in \cite{Crew:2020psc}. That work also provides an alternative expression for generic $F$, written in terms of the Milne polynomials. As discussed in \cite{Bullimore:2020jdq,Crew:2020psc}, this Hilbert series can be interpreted as a refined Verma character of the quantized Higgs branch algebra.
\\

For $F = 1$, the Higgs and Coulomb branches are identical since the theory is self-mirror. Thus, \eqref{eq:HS} also corresponds to the Coulomb branch Hilbert series of the ADHM with a single flavor, where $z$ is now associated with the topological $U(1)$ symmetry acting on monopole operators instead of the adjoint hyper. Conversely, one can derive the same Hilbert series in different expressions from the Coulomb branch side \cite{Hayashi:2022ldo,Hayashi:2024jof}. For instance, the following expression is obtained using the Fermi-gas method \cite{Hayashi:2022ldo}:
\begin{align}
\label{eq:HNO}
\mathcal H_{N;F=1}\left(z = (x_1/x_2)^\frac12,r = (x_1 x_2)^\frac12\right) = \sum_{|\lambda| = N} \prod_{i = 1}^r \frac{Z_{\lambda_i}[\mathbb C^4;N] (x_1;x_2)^{m_i}}{\lambda_i^{m_i} m_i!}
\end{align}
where
\begin{align}
Z_{i}[\mathbb C^4;N] (x_1;x_2) = \frac{1}{\left(1-x_1^i\right) \left(1-x_2^i\right)} \,.
\end{align}
We have confirmed the equality of the two expressions for several values of $N$. Note that \eqref{eq:HNO} also involves a Young diagram summation, but its interpretation differs from that in our formula: in our case, the Young diagrams are associated with discrete Higgs vacua, whereas in \eqref{eq:HNO} they are related to counting of monopole fluxes. This distinction reflects the self-mirror symmetry of the theory, which exchanges the Higgs and Coulomb branches. Additionally, it was also checked that \eqref{eq:HNO} can be re-expanded in terms of giant graviton contributions \cite{Hayashi:2024aaf}.

Moreover, one can take further limits of the Higgs branch Hilbert series:
\begin{align}
&r, \, z^{-1} \; \rightarrow \; 0, \qquad r z \text{ fixed}, \label{eq:half-BPS} \\
&r, \, z \quad \; \rightarrow \; 0, \qquad r/z \text{ fixed},
\end{align}
leading to the indices for half-BPS states. In each limit, only the contribution from the linear Young diagram $\tiny\begin{ytableau}[] & \\ \end{ytableau} \cdots \begin{ytableau}[] \\ \end{ytableau}^T$ or $\tiny\begin{ytableau}[] & \\ \end{ytableau} \cdots \begin{ytableau}[] \\ \end{ytableau}$ survives, respectively; namely,
\begin{align}
\lim_{\substack{r,z^{-1} \rightarrow 0 \\ s \equiv r z \text{ fixed}}} \mathcal H_{N;F=1} \quad &= \quad \left\{
\begin{matrix}
\prod_{i = 1}^N \frac{1}{1-s^i}\,, \qquad & \text{for } \tiny\begin{ytableau}[] & \\ \end{ytableau} \cdots \begin{ytableau}[] \\ \end{ytableau}^T \,, \\
0 \,, \quad & \text{otherwise} \,.
\end{matrix}
\right. \\
\lim_{\substack{r,z \rightarrow 0 \\ s \equiv r/z \text{ fixed}}} \mathcal H_{N;F=1} \quad &= \quad \left\{
\begin{matrix}
\prod_{i = 1}^N \frac{1}{1-s^i} \,, \qquad & \text{for } \tiny\begin{ytableau}[] & \\ \end{ytableau} \cdots \begin{ytableau}[] \\ \end{ytableau} \,, \\
0 \,, \quad & \text{otherwise} \,,
\end{matrix}
\right.
\end{align}
This is consistent with our expectation discussed in the previous section that the linear Young diagrams capture biased charge sectors, $Q_1 \gg Q_2$ and $Q_1 \ll Q_2$, as the states contributing these half-BPS indices satisfy
\begin{align}
E &= Q_1 \,, \qquad Q_2 = Q_3 = Q_4 = j = 0 \,, \\
E &= Q_2 \,, \qquad Q_1 = Q_3 = Q_4 = j = 0 \,,
\end{align}
respectively.
\\

For multiple flavors, on the other hand, the theory is no longer self-mirror; instead its mirror dual is given by a circular quiver of $F$ $U(N)$ gauge nodes one of which has a single flavor attached. Thus, \eqref{eq:HS} also corresponds to the Coulomb Hilbert series of the mirror circular quiver, where $\vec y$ and $z$ are now associated with FI parameters for the topological $U(1)$ symmetries rather than flavor real masses. As far as we are aware, \eqref{eq:HS} has not been derived in the literature, although there are alternative expressions \cite{Crew:2020psc,Hayashi:2024jof}, which involve the \emph{infinite} summations over Young diagrams of arbitrary size. In contrast, our expression restricts the contributing Young diagrams to the fixed total size $N$, providing a closed-form answer without infinite summations for a given $N$. It would be interesting to understand the analytic relation among these expressions.

One can also take the half-BPS limit, in which, again, only the linear Young diagrams contribute to the index. For instance, if we consider the limit \eqref{eq:half-BPS} for the $N = F = 2 $ case, the contributing Young diagrams are as follows:
\begin{align}
\begin{ytableau}[]
*(blue!67) \\
*(blue!67)
\end{ytableau}
&: \quad \frac{r y^{-4}}{\left(1-r\right) \left(1-r^2\right) \left(1-y^{-2}\right) \left(1-r y^{-2}\right)} \\
\begin{ytableau}[]
*(red!67) \\
*(red!67)
\end{ytableau}
&: \quad \frac{r y^4}{\left(1-r\right) \left(1-r^2\right) \left(1-y^2\right) \left(1-r y^2\right)} \\
\begin{ytableau}[]
*(blue!67)
\end{ytableau}
\quad
\begin{ytableau}[]
*(red!67)
\end{ytableau}
&: \quad \frac{1}{(1-r)^2 \left(1-r y^{-2}\right) \left(1-r y^2\right)}
\end{align}
The complete half-BPS index is given by
\begin{align}
\mathcal I^\text{half-BPS}_{N=2;F=2}(s) = \frac{1}{\left(1-s\right) \left(1-s^2\right)} \,,
\end{align}
or generally,
\begin{align}
\mathcal I^\text{half-BPS}_{N;F}(s) = \prod_{i = 1}^N \frac{1}{1-s^i} \,,
\end{align}
which reproduces the result of \cite{Hayashi:2024aaf}. We have confirmed this result for several values of $N$ and $F$. Note that the result after the summation is independent of $F$ and the flavor fugacities $y_i$.
\\

\section{Discussion}
\label{sec:discussion}
In this work, we have analyzed both the microcanonical and canonical aspects of the superconformal index for the 3d ADHM quiver gauge theory, which provides a UV description of the 3d $\mathcal N=8$ SCFT dual to M‑theory on AdS$_4 \times S^7$. By computing the index to sufficiently high orders using the factorized index, we probed the finite‑$N$ spectrum beyond the reach of the standard matrix integral formula. Our results reveal clear signatures of quantum black hole states: the finite‑$N$ degeneracies of the ADHM quiver exhibit remarkable agreement with the leading large‑$N$ contribution, which reproduces the holographic dual black hole entropy. It would be interesting to understand the characteristic behavior of quantum black hole states we observed---such as the alternating statistics and the growth of degeneracy---from the perspective of fortuitous states, which have recently been discussed in ABJ(M) theories with $k > 1$ \cite{Kim:2025vup,Belin:2025hsg,Behan:2025hbx}. In addition, our factorized index is written as a sum over massive Higgs vacua, with each summand given by the square of a holomorphic block. While we study each block numerically, it would be interesting to understand its asymptotic behavior analytically, in a manner analogous to, for example, a recent study of 4d indices \cite{Purkayastha:2025whi}.

From the canonical perspective, we introduced the complex‑$\beta$ phase diagram of the index, offering a new lens to visualize the contributions of distinct sets of states carrying different (complexified) entropies. This phase diagram exhibits multiple peaks, including subdominant structures that are suggestive of additional gravitational saddles beyond the standard graviton and black hole contributions.

We also discussed the generalization of the factorized index to the multi-flavor case, dual to M-theory on AdS$_4 \times S^7/\mathbb Z_F$, as well as its Higgs branch Hilbert series limit, which further illuminates the distribution of BPS states across discrete Higgs vacua, which remains invisible in the unrefined index.

Our study demonstrates that the finite‑$N$ superconformal index is a powerful probe of the microscopic black hole states, and more broadly, of quantum‑gravitational physics. Nevertheless, several limitations remain. First, our analysis of the complex-$\beta$ phase diagram was carried out only for the $N=2, \, 3$ cases. For higher $N$, we have obtained phase diagrams with qualitatively similar multiple peaks, but these rely on approximate indices computed using truncated vortex contributions. While such truncation reliably captures the dominant contributions as explained in Section~\ref{sec:SCI}, it may not accurately resolve the finer, subdominant peaks. For this reason, we postpone a complete presentation of the higher-$N$ phase diagrams until more precise index data are available.

Second, our analysis focused on the unrefined index, which does not retain refined charge information of the states contributing to each peak in the phase diagram. Extending our study to the refined index is computationally challenging, but highly worthwhile, as it would provide sharper insight into the origin of the peaks and clarify their correspondence to distinct gravitational saddles.

Along these lines, it would be particularly interesting to investigate how the second peak at nonzero $\theta$ observed in the $N=3$ case evolves for higher $N$, especially when refined charge information is included. In the unrefined index studied here, this peak remains subdominant for all values of $\mathrm{Re} \, \beta$, which is the only (real) chemical potential available to tune. By contrast, the refined index introduces additional chemical potentials, making it natural to ask whether there exists a charge sector in which this subdominant peak could become dominant. More generally, the connection between our \emph{finite}-$N$ phases and multiple phases from the generalized Cardy limit of the \emph{large}-$N$ superconformal index, discussed in \cite{BenettiGenolini:2023rkq}, is worth studying. Finally, it would be also interesting to analyze the indices of other models, such as 4d $\mathcal N=4$ SYM, using their complex-$\beta$ phase diagrams.
\\

\acknowledgments

We would like to thank Sung-Soo Kim and Kimyeong Lee for useful discussions. This work is supported by the National Natural Science Foundation of China under Grant No.~12247103.

\newpage
\appendix

\section{Numerical index data}
\label{sec:data}

\subsection{$F = 1$}

\begin{table}[h!]
\centering
\vspace{2cm}
\begin{tabular}{|c|c||c|c||c|c||c|c|}
\hline
$Q$ & $I(Q)$ & $Q$ & $I(Q)$ & $Q$ & $I(Q)$ & $Q$ & $I(Q)$ \\
\hline
1 & 4 & 26 & 508 & 51 & $-14440$ & 76 & $-43579$ \\
2 & 10 & 27 & 592 & 52 & $-1055$ & 77 & $-249000$ \\
3 & 16 & 28 & $-212$ & 53 & 18412 & 78 & $-5922$ \\
4 & 19 & 29 & $-664$ & 54 & 7024 & 79 & 302512 \\
5 & 20 & 30 & 264 & 55 & $-20864$ & 80 & 79221 \\
6 & 26 & 31 & 1288 & 56 & $-14355$ & 81 & $-348812$ \\
7 & 40 & 32 & 491 & 57 & 22808 & 82 & $-177254$ \\
8 & 49 & 33 & $-1196$ & 58 & 25120 & 83 & 382864 \\
9 & 40 & 34 & $-794$ & 59 & $-20952$ & 84 & 304431 \\
10 & 26 & 35 & 1592 & 60 & $-36570$ & 85 & $-392332$ \\
11 & 40 & 36 & 1940 & 61 & 16820 & 86 & $-456322$ \\
12 & 84 & 37 & $-1064$ & 62 & 50774 & 87 & 371544 \\
13 & 100 & 38 & $-2614$ & 63 & $-6872$ & 88 & 634735 \\
14 & 52 & 39 & 784 & 64 & $-64685$ & 89 & $-305904$ \\
15 & 8 & 40 & 4038 & 65 & $-8136$ & 90 & $-832436$ \\
16 & 64 & 41 & 772 & 66 & 78812 & 91 & 184840 \\
17 & 172 & 42 & $-4510$ & 67 & 31512 & 92 & 1042745 \\
18 & 150 & 43 & $-2072$ & 68 & $-88482$ & 93 & 7284 \\
19 & $-16$ & 44 & 5503 & 69 & $-60984$ & 94 & $-1248186$ \\
20 & $-61$ & 45 & 4800 & 70 & 94128 & 95 & $-276960$ \\
21 & 172 & 46 & $-5158$ & 71 & 99448 & 96 & 1436485 \\
22 & 376 & 47 & $-7328$ & 72 & $-90291$ & 97 & 638384 \\
23 & 152 & 48 & 4771 & 73 & $-144160$ & 98 & $-1582312$ \\
24 & $-235$ & 49 & 11284 & 74 & 75678 & 99 & $-1096384$ \\
25 & $-96$ & 50 & $-2126$ & 75 & 196072 & 100 & 1659947 \\
\hline
\end{tabular}
\caption{\label{tab:n=1} $I_{N=1}(Q)$.}
\end{table}

\newpage
\begin{table}[h!]
\centering
\vspace{2cm}
\begin{tabular}{|c|c||c|c||c|c||c|c|}
\hline
$Q$ & $I(Q)$ & $Q$ & $I(Q)$ & $Q$ & $I(Q)$ & $Q$ & $I(Q)$ \\
\hline
1 & 4 & 26 & 53512 & 51 & $-3039696$ & 76 & 253569860 \\
2 & 20 & 27 & 35192 & 52 & $-5774712$ & 77 & 436570264 \\
3 & 56 & 28 & 7529 & 53 & 2096984 & 78 & $-173515228$ \\
4 & 139 & 29 & 27116 & 54 & 10016312 & 79 & $-662909992$ \\
5 & 260 & 30 & 95420 & 55 & 2420056 & 80 & $-18800137$ \\
6 & 436 & 31 & 118456 & 56 & $-12720055$ & 81 & 897555608 \\
7 & 640 & 32 & 31737 & 57 & $-8899188$ & 82 & 370978500 \\
8 & 954 & 33 & $-51484$ & 58 & 14801988 & 83 & $-1075169544$ \\
9 & 1420 & 34 & 47252 & 59 & 20406648 & 84 & $-905873624$ \\
10 & 2076 & 35 & 255504 & 60 & $-10987758$ & 85 & 1117450928 \\
11 & 2720 & 36 & 241333 & 61 & $-33345668$ & 86 & 1641225992 \\
12 & 3234 & 37 & $-82896$ & 62 & 1342452 & 87 & $-898257984$ \\
13 & 3780 & 38 & $-239576$ & 63 & 48773112 & 88 & $-2536798887$ \\
14 & 5012 & 39 & 191560 & 64 & 20743090 & 89 & 285315372 \\
15 & 7048 & 40 & 702809 & 65 & $-58643244$ & 90 & 3510893952 \\
16 & 8969 & 41 & 344652 & 66 & $-53725940$ & 91 & 883764784 \\
17 & 9372 & 42 & $-630708$ & 67 & 60119072 & 92 & $-4379393839$ \\
18 & 9160 & 43 & $-608536$ & 68 & 102064545 & 93 & $-2734051912$ \\
19 & 11504 & 44 & 915744 & 69 & $-39414584$ & 94 & 4881456348 \\
20 & 17743 & 45 & 1736912 & 70 & $-157547640$ & 95 & 5359732192 \\
21 & 22788 & 46 & $-102540$ & 71 & $-9736832$ & 96 & $-4622751190$ \\
22 & 20236 & 47 & $-2297376$ & 72 & 215741446 & 97 & $-8722854996$ \\
23 & 14096 & 48 & $-669835$ & 73 & 103428652 & 98 & 3130376152 \\
24 & 19366 & 49 & 3480996 & 74 & $-254224568$ & 99 & 12637829184 \\
25 & 40104 & 50 & 3218668 & 75 & $-243562936$ & 100 & 168588591 \\
\hline
\end{tabular}
\caption{\label{tab:n=2} $I_{N=2}(Q)$.}
\end{table}

\newpage
\begin{table}[h!]
\centering
\vspace{2cm}
\begin{tabular}{|c|c||c|c||c|c||c|c|}
\hline
$Q$ & $I(Q)$ & $Q$ & $I(Q)$ & $Q$ & $I(Q)$ & $Q$ & $I(Q)$ \\
\hline
1 & 4 & 26 & 1686030 & 51 & $-159286180$ & 76 & $-62074824537$ \\
2 & 20 & 27 & 1909628 & 52 & $-308279461$ & 77 & 76102730060 \\
3 & 76 & 28 & 2638781 & 53 & 165220072 & 78 & 136344817966 \\
4 & 239 & 29 & 3730916 & 54 & 741621172 & 79 & $-40980922696$ \\
5 & 644 & 30 & 4313620 & 55 & 358313432 & 80 & $-226940640423$ \\
6 & 1512 & 31 & 3862944 & 56 & $-826171406$ & 81 & $-57666714244$ \\
7 & 3100 & 32 & 3597751 & 57 & $-959372832$ & 82 & 311772645310 \\
8 & 5743 & 33 & 5632696 & 58 & 895231242 & 83 & 250591469676 \\
9 & 9856 & 34 & 9545336 & 59 & 2313071704 & 84 & $-335087722422$ \\
10 & 16182 & 35 & 11012600 & 60 & 237084373 & 85 & $-545477313496$ \\
11 & 25988 & 36 & 7141251 & 61 & $-3323858932$ & 86 & 220959177358 \\
12 & 40764 & 37 & 3509216 & 62 & $-2192044918$ & 87 & 924071216868 \\
13 & 61252 & 38 & 10072680 & 63 & 4205208120 & 88 & 140650293476 \\
14 & 87066 & 39 & 24834016 & 64 & 6328259945 & 89 & $-1294542646004$ \\
15 & 118372 & 40 & 28225032 & 65 & $-2497800124$ & 90 & $-850960563252$ \\
16 & 159560 & 41 & 8349524 & 66 & $-11107330518$ & 91 & 1486961159404 \\
17 & 219668 & 42 & $-9181610$ & 67 & $-2303809220$ & 92 & 1983734619134 \\
18 & 304214 & 43 & 16446364 & 68 & 16404416584 & 93 & $-1209489319964$ \\
19 & 404012 & 44 & 71198813 & 69 & 13586908084 & 94 & $-3488673536444$ \\
20 & 501321 & 45 & 73190244 & 70 & $-17158447216$ & 95 & 85767262344 \\
21 & 601724 & 46 & $-13749410$ & 71 & $-30379175116$ & 96 & 5129595055044 \\
22 & 757308 & 47 & $-74428808$ & 72 & 9750606932 & 97 & 2321893736312 \\
23 & 1021096 & 48 & 36543082 & 73 & 52843060108 & 98 & $-6347858136498$ \\
24 & 1347744 & 49 & 225910956 & 74 & 15219775806 & 99 & $-6351325819584$ \\
25 & 1586892 & 50 & 180681696 & 75 & $-71782352304$ & 100 & 6216576038392 \\
\hline
\end{tabular}
\caption{\label{tab:n=3} $I_{N=3}(Q)$.}
\end{table}

\newpage
\begin{table}[h!]
\centering
\vspace{2cm}
\begin{tabular}{|c|c||c|c||c|c||c|c|}
\hline
$Q$ & $I(Q)$ & $Q$ & $I(Q)$ & $Q$ & $I(Q)$ & $Q$ & $I(Q)$ \\
\hline
1 & 4 & 26 & 36457560 & 51 & $-2436265004$ & 76 & $(-6223609696918)$ \\
2 & 20 & 27 & 48853180 & 52 & 881969440 & 77 & $(440150937208)$ \\
3 & 76 & 28 & 62836622 & 53 & 13697547476 & 78 & $(10770560439192)$ \\
4 & 274 & 29 & 76771608 & 54 & 19817672520 & 79 & $(6747229719880)$ \\
5 & 844 & 30 & 92954408 & 55 & 3014230384 & 80 & $(-13060343807655)$ \\
6 & 2392 & 31 & 118862728 & 56 & $-20636056503$ & 81 & $(-19683510863092)$ \\
7 & 6040 & 32 & 159603602 & 57 & $-8209785248$ & 82 & $(8526603268424)$ \\
8 & 13973 & 33 & 206382512 & 58 & 46366358632 & 83 & $(38051178045278)$ \\
9 & 29456 & 34 & 240071416 & 59 & 70563025880 & 84 & $(11266819067398)$ \\
10 & 57756 & 35 & 259924364 & 60 & $-4803998984$ & 85 & $(-54345275551784)$ \\
11 & 106244 & 36 & 307535466 & 61 & $-105051587688$ & 86 & $(-53166253002008)$ \\
12 & 187074 & 37 & 431664844 & 62 & $-48141869368$ & 87 & $(54116553589374)$ \\
13 & 318496 & 38 & 600954848 & 63 & 175601554672 & 88 & $(120668995097068)$ \\
14 & 528272 & 39 & 685943784 & 64 & 252380923892 & 89 & $(-10369840695414)$ \\
15 & 850008 & 40 & 620664946 & 65 & $-75030764316$ & 90 & $(-200009550126532)$ \\
16 & 1320718 & 41 & 602804664 & 66 & $-452431986000$ & 91 & $(-108042462665298)$ \\
17 & 1978348 & 42 & 965846444 & 67 & $-165080317008$ & 92 & $(253014357573104)$ \\
18 & 2883596 & 43 & 1650608792 & 68 & $(704404668000)$ & 93 & $(328870601862792)$ \\
19 & 4148536 & 44 & 1948412418 & 69 & $(858567144108)$ & 94 & $(-200357949547956)$ \\
20 & 5949400 & 45 & 1302263040 & 70 & $(-509962889032)$ & 95 & $(-643520918943108)$ \\
21 & 8453160 & 46 & 568794608 & 71 & $(-1759273627320)$ & 96 & $(-71896254788310)$ \\
22 & 11716184 & 47 & 1628052208 & 72 & $(-326303271871)$ & 97 & $(973940843911658)$ \\
23 & 15697320 & 48 & 4575915690 & 73 & $(2819208970726)$ & 98 & $(696522596278186)$ \\
24 & 20578417 & 49 & 5925821840 & 74 & $(2636463393252)$ & 99 & $(-1116983116530986)$ \\
25 & 27124976 & 50 & 2374288436 & 75 & $(-2772404318490)$ & 100 & $(-1752536951200026)$ \\
\hline
\end{tabular}
\caption{\label{tab:n=4} $I_{N=4}(Q)$ obtained by keeping vortex contributions up to vorticity 67. The values for $Q \leq 67$ are exact, whereas those for $Q > 67$ are computed approximately due to truncated vortex contributions.}
\end{table}

\newpage
\begin{table}[h!]
\centering
\vspace{2cm}
\begin{tabular}{|c|c||c|c||c|c||c|c|}
\hline
$Q$ & $I(Q)$ & $Q$ & $I(Q)$ & $Q$ & $I(Q)$ & $Q$ & $I(Q)$ \\
\hline
1 & 4 & 26 & 350266614 & 51 & $(113358021050)$ & 76 & $(-170645601418566)$ \\
2 & 20 & 27 & 483346216 & 52 & $(225451238050)$ & 77 & $(-694265104756)$ \\
3 & 76 & 28 & 663109157 & 53 & $(310570543402)$ & 78 & $(315640337778406)$ \\
4 & 274 & 29 & 911336272 & 54 & $(231049497994)$ & 79 & $(272683959004882)$ \\
5 & 900 & 30 & 1247902572 & 55 & $(45168594768)$ & 80 & $(-328246800113430)$ \\
6 & 2742 & 31 & 1677024564 & 56 & $(89793256744)$ & 81 & $(-740290896259610)$ \\
7 & 7720 & 32 & 2191866317 & 57 & $(589854003244)$ & 82 & $(-2408722459444)$ \\
8 & 20218 & 33 & 2815773580 & 58 & $(1103040590064)$ & 83 & $(1324774027470558)$ \\
9 & 49236 & 34 & 3649448462 & 59 & $(750528055608)$ & 84 & $(1080373189587376)$ \\
10 & 112018 & 35 & 4838036988 & 60 & $(-456061230074)$ & 85 & $(-1457715507653672)$ \\
11 & 238924 & 36 & 6417691493 & 61 & $(-786755064470)$ & 86 & $(-3028964475617002)$ \\
12 & 481019 & 37 & 8193585416 & 62 & $(1394460860872)$ & 87 & $(205167642457532)$ \\
13 & 920992 & 38 & 9940767856 & 63 & $(4294700538546)$ & 88 & $(5485039399759195)$ \\
14 & 1692246 & 39 & 11942144420 & 64 & $(3083764505338)$ & 89 & $(3909870278320838)$ \\
15 & 3007540 & 40 & 15207255260 & 65 & $(-3291613780886)$ & 90 & $(-6523005137450692)$ \\
16 & 5198335 & 41 & 20530721688 & 66 & $(-6585369988742)$ & 91 & $(-11736890493314968)$ \\
17 & 8752344 & 42 & 26862112292 & 67 & $(2730927754156)$ & 92 & $(2450463341163749)$ \\
18 & 14340048 & 43 & 31447146968 & 68 & $(17686898075237)$ & 93 & $(22162517646188370)$ \\
19 & 22837992 & 44 & 33588894373 & 69 & $(14103455632828)$ & 94 & $(12371897841071696)$ \\
20 & 35414847 & 45 & 38728314996 & 70 & $(-16425609972754)$ & 95 & $(-28840217071078398)$ \\
21 & 53759388 & 46 & 54987723746 & 71 & $(-36166437364526)$ & 96 & $(-42416672830433973)$ \\
22 & 80431652 & 47 & 80471364948 & 72 & $(3447458674412)$ & 97 & $(18105147531830566)$ \\
23 & 119056196 & 48 & 96426330751 & 73 & $(74601219250310)$ & 98 & $(85406423169118680)$ \\
24 & 173985404 & 49 & 86624550748 & 74 & $(63873446706520)$ & 99 & $(31653130184236946)$ \\
25 & 249562588 & 50 & 72964896878 & 75 & $(-74230558747910)$ & 100 & $(-121177080926622458)$ \\
\hline
\end{tabular}
\caption{\label{tab:n=5} $I_{N=5}(Q)$ obtained by keeping vortex contributions up to vorticity 50. The values for $Q \leq 50$ are exact, whereas those for $Q > 50$ are computed approximately due to truncated vortex contributions.}
\end{table}

\newpage
\subsection{$F=2$}

\begin{table}[h!]
\centering
\vspace{2cm}
\begin{tabular}{|c|c||c|c||c|c||c|c|}
\hline
$Q$ & $I(Q)$ & $Q$ & $I(Q)$ & $Q$ & $I(Q)$ & $Q$ & $I(Q)$ \\
\hline
1 & 2 & 26 & $-1667$ & 51 & 138078 & 76 & $-1338068$ \\
2 & 9 & 27 & 494 & 52 & $-9664$ & 77 & 4131852 \\
3 & 14 & 28 & 3051 & 53 & $-187646$ & 78 & 2963096 \\
4 & 22 & 29 & 1228 & 54 & $-56129$ & 79 & $-4390550$ \\
5 & 20 & 30 & $-3537$ & 55 & 230122 & 80 & $-5085734$ \\
6 & 25 & 31 & $-3284$ & 56 & 158687 & 81 & 4093812 \\
7 & 34 & 32 & 3927 & 57 & $-247648$ & 82 & 7674449 \\
8 & 62 & 33 & 7182 & 58 & $-298296$ & 83 & $-2966220$ \\
9 & 74 & 34 & $-1517$ & 59 & 222776 & 84 & $-10603382$ \\
10 & 46 & 35 & $-10500$ & 60 & 472972 & 85 & 709088 \\
11 & 2 & 36 & $-2324$ & 61 & $-128144$ & 86 & 13631274 \\
12 & 38 & 37 & 14354 & 62 & $-663423$ & 87 & 2980470 \\
13 & 148 & 38 & 10279 & 63 & $-56104$ & 88 & $-16368176$ \\
14 & 213 & 39 & $-15090$ & 64 & 850307 & 89 & $-8348734$ \\
15 & 76 & 40 & $-20788$ & 65 & 357528 & 90 & 18294046 \\
16 & $-108$ & 41 & 12306 & 66 & $-991498$ & 91 & 15573790 \\
17 & $-24$ & 42 & 34968 & 67 & $-790238$ & 92 & $-18708245$ \\
18 & 353 & 43 & $-1348$ & 68 & 1033051 & 93 & $-24674408$ \\
19 & 494 & 44 & $-47820$ & 69 & 1360464 & 94 & 16746813 \\
20 & $-17$ & 45 & $-17358$ & 70 & $-900663$ & 95 & 35440914 \\
21 & $-562$ & 46 & 58752 & 71 & $-2044600$ & 96 & $-11413509$ \\
22 & $-62$ & 47 & 48122 & 72 & 520535 & 97 & $-47342460$ \\
23 & 1098 & 48 & $-60190$ & 73 & 2800420 & 98 & 1665717 \\
24 & 1082 & 49 & $-88378$ & 74 & 198685 & 99 & 59485744 \\
25 & $-714$ & 50 & 47435 & 75 & $-3539082$ & 100 & 13558479 \\
\hline
\end{tabular}
\caption{\label{tab:n=1-f=2} $I_{N=1; \, F=2}(Q)$.}
\end{table}

\newpage
\begin{table}[h!]
\centering
\vspace{2cm}
\begin{tabular}{|c|c||c|c||c|c||c|c|}
\hline
$Q$ & $I(Q)$ & $Q$ & $I(Q)$ & $Q$ & $I(Q)$ & $Q$ & $I(Q)$ \\
\hline
1 & 2 & 26 & $-15266$ & 51 & $-28214160$ & 76 & 12730004128 \\
2 & 12 & 27 & 89760 & 52 & 61644066 & 77 & 1700777478 \\
3 & 32 & 28 & 214210 & 53 & 84398696 & 78 & $-17865168076$ \\
4 & 91 & 29 & 146002 & 54 & $-44252956$ & 79 & $-11318144748$ \\
5 & 188 & 30 & $-127564$ & 55 & $-159090032$ & 80 & 20596465611 \\
6 & 368 & 31 & $-226788$ & 56 & $-25833842$ & 81 & 26555242838 \\
7 & 564 & 32 & 200596 & 57 & 234626916 & 82 & $-17167902542$ \\
8 & 854 & 33 & 755096 & 58 & 180963038 & 83 & $-46832954924$ \\
9 & 1208 & 34 & 471302 & 59 & $-256616892$ & 84 & 2596389689 \\
10 & 1864 & 35 & $-715756$ & 60 & $-427916668$ & 85 & 69017948570 \\
11 & 2844 & 36 & $-1141120$ & 61 & 150876730 & 86 & 28673929778 \\
12 & 4058 & 37 & 614030 & 62 & 745701122 & 87 & $-86101815120$ \\
13 & 4838 & 38 & 2728480 & 63 & 192031300 & 88 & $-81132917251$ \\
14 & 5140 & 39 & 1350944 & 64 & $-1029986660$ & 89 & 86284866690 \\
15 & 5900 & 40 & $-3239544$ & 65 & $-862244208$ & 90 & 155704165064 \\
16 & 9171 & 41 & $-4241858$ & 66 & 1091715572 & 91 & $-52688258488$ \\
17 & 14714 & 42 & 2791982 & 67 & 1902555920 & 92 & $-246034118886$ \\
18 & 18594 & 43 & 9508748 & 68 & $-617476497$ & 93 & $-35212471212$ \\
19 & 15524 & 44 & 2547134 & 69 & $-3191678700$ & 94 & 334252481692 \\
20 & 9598 & 45 & $-13354066$ & 70 & $-765525846$ & 95 & 197951788116 \\
21 & 14984 & 46 & $-12706930$ & 71 & 4376648244 & 96 & $-386636616504$ \\
22 & 39832 & 47 & 13840208 & 72 & 3430829013 & 97 & $-449083501630$ \\
23 & 62304 & 48 & 30513476 & 73 & $-4736102218$ & 98 & 350962381742 \\
24 & 44434 & 49 & $-1573688$ & 74 & $-7524928328$ & 99 & 784818996808 \\
25 & $-8384$ & 50 & $-49287570$ & 75 & 3182379332 & 100 & $-157103700994$ \\
\hline
\end{tabular}
\caption{\label{tab:n=2-f=2} $I_{N=2; \, F=2}(Q)$.}
\end{table}

\newpage
\begin{table}[h!]
\centering
\vspace{2cm}
\begin{tabular}{|c|c||c|c||c|c||c|c|}
\hline
$Q$ & $I(Q)$ & $Q$ & $I(Q)$ & $Q$ & $I(Q)$ & $Q$ & $I(Q)$ \\
\hline
1 & 2 & 26 & 2708084 & 51 & $-3308754064$ & 76 & $-2699508634343$ \\
2 & 12 & 27 & 3907564 & 52 & $-230049973$ & 77 & 1537639946784 \\
3 & 36 & 28 & 4140077 & 53 & 6029530202 & 78 & 5512050125397 \\
4 & 118 & 29 & 2948354 & 54 & 5845459260 & 79 & 1215746735238 \\
5 & 312 & 30 & 2341876 & 55 & $-5618994068$ & 80 & $-8226198086744$ \\
6 & 813 & 31 & 5802064 & 56 & $-14913791837$ & 81 & $-7491811260068$ \\
7 & 1790 & 32 & 13160275 & 57 & $-2066770584$ & 82 & 8454529631001 \\
8 & 3719 & 33 & 16702212 & 58 & 25138228634 & 83 & 17770846319294 \\
9 & 6848 & 34 & 7968996 & 59 & 24520466528 & 84 & $-2092892125779$ \\
10 & 11892 & 35 & $-6294156$ & 60 & $-24057200450$ & 85 & $-30128441163626$ \\
11 & 19284 & 36 & $-752625$ & 61 & $-61695234172$ & 86 & $-16051554628012$ \\
12 & 30946 & 37 & 38298200 & 62 & $-6198559148$ & 87 & 38248437191214 \\
13 & 48948 & 38 & 73548473 & 63 & 102980876558 & 88 & 50120716717677 \\
14 & 77036 & 39 & 37556394 & 64 & 91065106796 & 89 & $-29628592287948$ \\
15 & 115898 & 40 & $-64433556$ & 65 & $-107147935712$ & 90 & $-98709223017124$ \\
16 & 163979 & 41 & $-91665848$ & 66 & $-237096548172$ & 91 & $-14265605578870$ \\
17 & 217240 & 42 & 87814356 & 67 & 5916314192 & 92 & 148211427764085 \\
18 & 283604 & 43 & 327519828 & 68 & 409752248143 & 93 & 113806793903852 \\
19 & 387832 & 44 & 225162299 & 69 & 293533792096 & 94 & $-165141902737509$ \\
20 & 563242 & 45 & $-323563792$ & 70 & $-478186311146$ & 95 & $-280065359191784$ \\
21 & 799184 & 46 & $-640582725$ & 71 & $-841397886796$ & 96 & 91352261495653 \\
22 & 1017841 & 47 & 112655698 & 72 & 200480758151 & 97 & 495272146154922 \\
23 & 1133038 & 48 & 1422324777 & 73 & 1555322957980 & 98 & 151741714454716 \\
24 & 1239764 & 49 & 1228364628 & 74 & 766396517490 & 99 & $-685731332901192$ \\
25 & 1675748 & 50 & $-1366691242$ & 75 & $-2052489446980$ & 100 & $-640733064860587$ \\
\hline
\end{tabular}
\caption{\label{tab:n=3-f=2} $I_{N=3; \, F=2}(Q)$.}
\end{table}

\newpage
\bibliographystyle{JHEP}
\bibliography{refs}
\end{document}